\documentclass[showpacs,preprintnumbers,amsmath,amssymb]{revtex4}

\usepackage{color}   
\usepackage{graphicx}
\usepackage{dcolumn}
\usepackage{bm}

\begin{document}
\def\bbox#1{\hbox{\boldmath${#1}$}}
\def\blambda{{\hbox{\boldmath $\lambda$}}}
\def\eeta{{\hbox{\boldmath $\eta$}}}
\def\bxi{{\hbox{\boldmath $\xi$}}}

\title{ Heavy Quarkonia in Quark-Gluon Plasma}

\author{Cheuk-Yin Wong}

\affiliation{Physics Division, Oak Ridge National Laboratory, Oak Ridge, TN
37831}

\affiliation{Department of Physics, University of Tennessee, Knoxville, TN 
37996}
\date{\today}

\begin{abstract}
Using the color-singlet free energy $F_1$ and total internal energy
$U_1$ obtained by Kaczmarek $et~al.$ for a static quark $Q$ and an
antiquark $\bar Q$ in quenched QCD, we study the binding energies and
wave functions of heavy quarkonia in a quark-gluon plasma. By
minimizing the grand potential in a simplified schematic model, we
find that the proper color-singlet $Q$-$\bar Q$ potential can be
obtained from the total internal energy $U_1$ by subtracting the gluon
internal energy contributions.  We carry out this subtraction in the
local energy-density approximation in which the gluon energy density
can be related to the local gluon pressure by the quark-gluon plasma
equation of state.  We find in this approximation that the proper
color-singlet $Q$-$\bar Q$ potential is approximately $F_1$ for $T\sim
T_c$ and it changes to $\frac{3}{4}F_1+\frac{1}{4}U_1$ at high
temperatures.  In this potential model, the $J/\psi$ is weakly bound above
the phase transition temperature $T_c$, and it dissociates
spontaneously above $1.62 T_c$, while $\chi_c$ and $\psi'$ are unbound
in the quark-gluon plasma.  The bottomium states $\Upsilon$, $\chi_b$
and $\Upsilon'$ are bound in the quark-gluon plasma and they
dissociate at $4.10 T_c, 1.18 T_c,$ and $1.38 T_c$ respectively.  For
comparison, we evaluate the heavy quarkonium binding energies also in
other models using the free energy $F_1$ or the total internal energy
$U_1$ as the $Q$-$\bar Q$ potential.  The comparison shows that the
model with the new $Q$-$\bar Q$ potential proposed in this manuscript
gives dissociation temperatures that agree best with those from
spectral function analyses.  We evaluate the cross section for
$\sigma(g+ J/\psi \to c + \bar c)$ and its inverse process, in order
to determine the $J/\psi$ dissociation width and the rate of $J/\psi$
production by recombining $c$ and $\bar c$ in the quark gluon plasma.

\end{abstract}

                                                                         
\maketitle
                                                                                
\section{Introduction}

The stability of heavy quarkonia in the quark-gluon plasma is an
interesting subject of current research in high-energy heavy-ion
collisions as Matsui and Satz suggested that the suppression of
$J/\psi$ production can be used as a signature of the quark-gluon
plasma \cite{Mat86}.  DeTar \cite{Det85,Det88}, Hansson, Lee, and
Zahed \cite{Han88}, and Simonov \cite{Sim95,Sim95a,Sim05} argued
however that because the range of strong interaction is not likely to
change drastically across the phase transition, low-lying mesons
including $J/\psi$ may remain relatively narrow states and the
suppression of $J/\psi$ is not a signature of the deconfinement phase
transition \cite{Det88}.  Whether or not $J/\psi$ production will be
suppressed depends on the screening between the heavy quark $Q$ and
the heavy antiquark $\bar Q$ when the quarkonium is placed in the
quark-gluon plasma.  The degree of screening is highly nonperturbative
at temperatures near the phase transition temperature
\cite{Kar03}. The related question of the quarkonium stability must be
examined in nonperturbative QCD using, for example, the lattice gauge
theory.
 
Recent investigations of masses and widths of heavy quarkonia in
quenched lattice QCD calculations were carried out by Asakawa $et~al.$
\cite{Asa03,Asa04} and Petreczky $et~al.$ \cite{Pet03,Dat03,Pet04}
using the spectral function analysis and the maximum entropy method.
They found that the width of $J/\psi$ remains relatively narrow up to
1.6 times the critical phase transition temperature $T_c$.
Reconsidering the properties of the quark-gluon plasma also led Zahed
and Shuryak to suggest that quark-gluon plasma at temperatures up to a
few $T_c$ supports weakly bound meson states
\cite{Shu03,Shu04a,Shu04b,Shu04c}. They have also estimated the
binding energy of $J/\psi$ and found it to be stable up to 2.7 $T_c$
\cite{Shu04a}.  The possibility of weakly bound meson states in the
quark-gluon plasma was suggested earlier by DeTar \cite{Det85,Det88},
Hatsuda, and Kunihiro \cite{Hat85}.  Phenomenological discussions on
medium modifications of charmonium in high-energy heavy-ion collision
have been presented recently by Grandchamp, Rapp, and Brown
\cite{Gra04}.  Summaries of recent development in heavy quarkonium
suppression and deconfinement have also been reported by Petreczky
\cite{Pet05} and Karsch \cite{Kar05}.

As the knowledge of the stability of $J/\psi$ has important
implications on the fate of $J/\psi$ in the quark-gluon plasma, it is
important to obtain an independent assessment on the binding of heavy
quarkonia, in addition to those from previous analyses.  The spectral
function analyses of heavy quarkonia using gauge-invariant
current-current correlators have been carried out in the quenched
approximation.  Within the quenched approximation, independent lattice
gauge calculations have also been carried out using the correlation of
Polyakov lines from which the free energy $F_1$ and the total internal
energy $U_1$ can be calculated
\cite{Kac03}.  The two-body potential obtained from the lattice gauge
theory can be used to study the dissociation of heavy quarkonia.  It
is of interest to ask whether, within the same quenched approximation,
the spectral function analysis and the potential model analysis will lead
to consistent results concerning the stability of quarkonia in the
quark-gluon plasma.  As we shall deal with lattice results from
quenched QCD only, the quark-gluon plasma we shall consider consists
of gluons.  For convenience, we shall continue to refer to such a
gluon medium from quenched QCD as a quark-gluon plasma.

Besides checking the consistency of independent quenched lattice gauge
calculations, we would like to use the potential model to examine many
physical quantities of interest.  If quarkonia are indeed stable in
the plasma, it is useful to find out how strongly bound they are.
Furthermore, $J/\psi$ can dissociate by collision with constituents of
the plasma.  In such a collisional dissociation, the rate of
dissociation depends on the cross section for the reaction $g+J/\psi
\to c + \bar c$.  We would like to evaluate this cross section as a
function of temperature $T$, which can be obtained by using the bound
state wave functions in the potential model.  The knowledge of the
dissociation cross section allows a determination of the collisional
dissociation width.  

In energetic heavy-ion collisions, many pairs of charm quarks and
antiquarks may be produced in a single central collision
\cite{The01a,The01b}.  These charm quarks and antiquarks can recombine
to form $J/\psi$ in the quark-gluon plasma.  We also wish to find out
here the rate of producing $J/\psi$ through such a reaction. The
production rate depends on the cross section for the reaction $c+\bar
c \to J/\psi + g$.  The latter quantity can be obtained from the cross
section for the inverse reaction $g+J/\psi \to c + \bar c$, which we
already intend to calculate.

Previously, the effects of temperature on the stability of heavy
quarkonium was studied by Digal $et~al.$\cite{Dig01a,Dig01b} and Wong
\cite{Won01a,Won02} using the free energy and assuming that the
effects of entropy are small.  It was, however, pointed out by Zantow,
Kaczmarek, Karsch and Petreczky \cite{Kac02,Kar03,Zan03} that the
effects of entropy depend on the separation distance between $c$ and
$\bar c$.  They suggested that the total internal energy $U_{1}$,
instead of the free energy $F_1$, may be used as the $Q\bar Q$
potential for the calculation of heavy quarkonium bound states. As the
theoretical basis for this suggestion has not been fully explained in
the literature, we shall go into details to examine the theoretical
questions on the proper potential for $Q$-$\bar Q$ states.  We find
that the proper $Q$-$\bar Q$ potential involves the $Q$ and $\bar
Q$ internal energy $U_{Q\bar Q}^{(1)}$.  We shall show that in the
local energy-density approximation, $U_{Q\bar
Q}^{(1)}=3F_1/(3+a)+aU_1/(3+a)$ where $a=3p/\epsilon$ is given by the
quark-gluon plasma equation of state.

When a heavy quarkonium is placed in a quark-gluon plasma,
conventional description assumes that the medium effect is dominated
by the effect of Debye screening \cite{Mat86,Kar03}, which leads to a
decrease in the attractive interaction between the heavy quark and
antiquark. We would like to study the effects of antiscreening due to
the deconfined gluons and the relationship between antiscreening and
the area law of spatial Polyakov loops
\cite{Sve82,Yaf82,Det85,Deg86,Det88}.  We would like to show that
because the Gauss law of QCD contains a non-linear term involving the
gluon field, the gluon field induces color charges at the field
points.  These induced color charges act to antiscreen the interaction
between the heavy quark and the antiquark.  We shall show that the
strength of the antiscreening effect increases with an increase in the
gluon correlation length and is proportional quadratically to the
magnitude of the gluon fields. The antiscreening effects due to
deconfined gluons bring an additional degree of freedom to mediate the
interaction between the quark and the antiquark.

This paper is organized as follows.  In Section II, we review the
heavy quarkonium production mechanism and the thermalization of the
quark-gluon medium in high-energy heavy-ion collisions.  We examine
the evidence for rapid thermalization as revealed by the elliptic flow
and hydrodynamics.  In Section III, we review the lattice gauge
calculations for the interaction between a heavy quark and a heavy
antiquark and the gauge dependence of the interaction in bound state
problems.  In Section IV, we show that the total internal energy
$U_{1}$ contains contributions from the internal energy of the
$Q$-$\bar Q$ pair and the internal energy of the gluons.  In Section
V, we use an appropriate variational principle to obtain the equation
of motion for the quarkonium single-particle states and find that the
proper $Q$-$\bar Q$ potential involves only the $Q$-$\bar Q$ internal
energy.  In order to obtain the internal energy of the heavy quark
pair, it is necessary to subtract the gluon internal energy from the
total internal energy $U_1$. In Section VI we show how such a
subtraction can be carried out in the local energy-density
approximation, using the quark-gluon plasma equation of state and the
First Law of Thermodynamics.  In Section VII, we show how the
color-singlet $F_1$ and $U_1$ obtained by Kaczmarek $et~al.$
\cite{Kac03} in quenched QCD can be parametrized and the proper
$Q$-$\bar Q$ potential can be obtained as a linear combination of
$F_1$ and $U_1$ from the lattice gauge results.  Using this heavy
quark-antiquark potential, we calculate the eigenenergies and
eigenfunctions for charmonia in the quark-gluon plasma as a function
of temperature in Section VIII.  The locations of the dissociation
temperatures at which heavy quarkonia begin to be unbound are then
determined. The heavy quarkonium dissociation temperatures are
compared with those determined from spectral function analyses. In
Section IX, we calculate the eigenenergies and eigenfunctions for $b
\bar b$ bound states in the quark-gluon plasma as a function of
temperature. We discuss the effects of antiscreening due to deconfined
gluons in the quark-gluon plasma in Section X.  In Section XI, we
discuss how the $J/\psi$ bound state wave function can be used to
calculate the cross section for $g+J/\psi \to c + \bar c$ after the
$J/\psi$ absorbs an E1 gluon, using the formulation of gluon
dissociation cross section presented previously \cite{Won02}.  The
dissociation cross sections and collisional dissociation widths of
$J/\psi$ in the quark-gluon plasma are then determined as a function
of temperature in Section XII. In Section XIII we evaluate the cross
section for the inverse process of $c +\bar c \to J/\psi +g$ using the
cross section of $g+J/\psi \to c + \bar c$ obtained in Section XII.
The rate of $J/\psi$ production by recombining $c$ and $\bar c$ in a
quark-gluon plasma is estimated.  We conclude our discussions in
Section XIV.  In Appendix A, we show that the integral of the gauge
fields along a space-like Polyakov loop obeys an area law if the gauge
fields are correlated.  This result is used in Section X to explain
the antiscreening effect.

\section{Heavy quarkonia production and the thermalization of the medium}

We are interested in using a heavy quarkonium to probe the properties
of the matter produced in central high-energy heavy-ion collisions.
In the collider frame, the colliding nuclei have the shape of
Lorentz-contracted disks.  The collisions are known to be highly
inelastic in which a large fraction of the incident collision energy
is released after the collision.  What is the rate of the relaxation
of the initial configuration to thermal equilibrium?

From the experimental viewpoint, recent RHIC experiments by the STAR
\cite{Sta01}, PHENIX \cite{Phe02}, and PHOBOS \cite{Pho02}
Collaborations reveal the presence of an elliptic collective flow in
non-central Au-Au collisions at RHIC energies.  The occurrence of such
a flow indicates that the initial azimuthally symmetric momentum
distribution of particles is deformed into an azimuthally asymmetric
momentum distribution.  The magnitude of the azimuthally asymmetry is
sensitive to the time at which the free streaming of particles
terminates and the dynamics of a thermally equilibrated system begins
\cite{Kol99,Hir01,Tea01}.  Too late a thermalization time will lead to
a spatially more extended system with a lower pressure gradient and a
smaller azimuthal asymmetry. The azimuthal asymmetry is also sensitive
to the numbers of degrees of freedom in the equation of state. The
magnitude of the asymmetry can be well explained in terms of a
hydrodynamical model of the quark-gluon plasma by assuming
thermalization at an initial time about 0.6 fm/c \cite{Kol99}.  We
infer from the experimental elliptic flow data and its hydrodynamical
description that the thermalization in the central region of a RHIC
nucleus-nucleus collision is very rapid, as short as 0.6 fm/c after
the collision.

From theoretical viewpoints, it was first pointed out by Landau
\cite{Lan54} that the initial configuration after a high-energy
nuclear collision consists of matter at an extremely high energy
density in a very thin disk.  The great magnitude of the energy
density means that the number density of quanta of matter is very
large.  Such a large number density in the thin disk of matter leads
to a very small mean-free-path compared to its dimensions, leading to
a rapid relaxation to thermal equilibrium.  According to Landau, ``in
the course of time, the system expands, the property of the small
mean-free path must be valid also for a significant part of the
process of expansion and this part of the expansion process must have
hydrodynamical character" \cite{Lan54}.  Landau hydrodynamics provides
a reasonable description of the widths of the rapidity distribution
for high-energy hadron-hadron and nucleus-nucleus collisions from
$\sqrt{s}=3$ GeV to RHIC collisions at 200 GeV
\cite{Car72,Mur04}.  Hydrodynamical description with a rapid thermal
relaxation also provides a good description of the elliptic flow of
matter after a Au-Au collision at RHIC, as indicated above.

It is also useful to point out that a quanta in the non-equilibrium
QCD matter interacts not only with other quanta (gluons and quarks) in
two-body processes in terms of two-body collisions, but the quanta
also interact with the fields generated by all other quanta. Because
of the non-Abelian nature of the QCD interaction, the fields generated
by other quanta are also sources of color fields and the quanta must
in addition interact with the color fields generated by the fields of
all other particles, in a highly non-linear manner (see Section X for
another manifestation of the non-linear nature of the gauge field).
Thus, a quanta interacts with other quanta not only by direct
short-range two-body collisions, but also by highly non-local
action-at-a-distance long-range interactions, through the fields
generated by the fields of other quanta.  There is thus an additional
non-linear and long-range mechanism of thermalization in non-Abelian
interactions which provides an extra push for rapid relaxation to
thermal equilibrium.

The rate of thermalization of a quark-gluon system after an
ultra-relativistic heavy-ion collision is the subject of current
theoretical research and has been discussed by Wong \cite{Won04},
Molnar, and Gyulassy \cite{Mol02}.  The small mean-free-path has also
been discussed by Shuryak \cite{Shu04}, Gyulassy and McLerran
\cite{Gyu04}, (see also Ref.\ \cite{Rik04}).  
The focus of the research is on trying to understand the
phenomenologically fast rate of thermalization as indicated be the
experimental elliptic flow evidence.  For example, in the work of
Molnar and Gyulassy \cite{Mol02}, it was necessary to shorten the
parton mean-free-path by an order of magnitude in order to reproduce
the magnitude of the elliptic flow.  Similarly, in the work of Lin,
Ko, and Pal \cite{Lin02} in parton cascade, it was necessary to
increase the parton-parton cross section by a large factor to describe
the dynamics in a nucleus-nucleus collisions at RHIC.

To use heavy quarkonia as a probe of the quark-gluon plasma, we need a
knowledge of the heavy-quarkonium production mechanism.  In a
nucleus-nucleus collision at high energies, the partons of one nucleon
and the partons of another nucleon can collide to produce occasionally
a pair of heavy quark and antiquark.  The time scale for the
production is of the order of $\hbar/2m_Q$ where $m_Q$ is the mass of
the heavy quark.  It is of order 0.06 fm/c for a $c$-$\bar c$ pair and
of order 0.02 fm/c for a $b$-$\bar b$ pair.  As the initial partons
carry varying fractions of the initial momenta of the colliding
nucleons, the heavy quark pair will come out at different energies.
Depending on the Feynman diagram of the production process, the
produced $Q$-$\bar Q$ pair after the hard scattering process may be in
a purely color-singlet quarkonia state or a coherent admixture of
color singlet- and color-octet states \cite{Bod95,Won99}.  The
projection of different final states from a coherent admixture gives
the probability amplitude for the occurrence of the final states.  A
color-octet state will need to emit a soft gluon of energy $E_{\rm
gluon}$ to become subsequently a color-singlet state in an emission
time of order $\hbar/ E_{\rm gluon}$.  For the emission of soft gluon
of a few hundred MeV, the time for the emission is of order 0.5-1.0
fm/c.

From the above considerations on the rapid thermalization of the
quark-gluon plasma and the time for the production of heavy quarkonia,
one envisages that by the time when the colliding matter is
thermalized at about 0.6 fm/c, a large fraction of the quarkonia have
already been formed, although in various energy states. The
quark-gluon plasma is expected to have a life time of a few fm/c which
is longer than the heavy quarkonium orbital period of order
$\hbar/(0.5$ GeV).  It is therefore meaningful to study the fate of a
produced heavy quarkonia in a thermalized quark-gluon plasma at a
finite temperature.  The behavior of the heavy quarkonium system
before the thermal equilibrium of the quark-gluon plasma and the
interaction of a coherent $Q$-$\bar Q$ color admixture in the
thermalized quark-gluon plasma are other topics which are beyond the
scope of the investigation of the present manuscript.

It should be pointed out that for a heavy quarkonium system in a
quark-gluon plasma the thermalization of the quark-gluon medium does
not necessarily imply the thermalization of the heavy quarkonium
system.  The former arises from the interaction among the light quarks
and gluons, while the latter depends on the interaction between the
heavy quarkonium and the constituents the quark-gluon plasma.  Our
evidence concerning the the rapid thermalization given above refers to
the thermalization of the quark-gluon plasma and not necessarily the
thermalization of the heavy quarkonium system in the quark-gluon
plasma.  If an isolated heavy quarkonium is placed in the thermalized
quark-gluon plasma, the heavy quarkonium system is not in thermal
equilibrium with the medium.  It will interact with the medium as its
density matrix will evolve with time.  Given a sufficient time that is
longer than the heavy quarkonium thermalization time, the heavy
quarkonium will also reach thermal equilibrium with its thermalized
quark-gluon plasma.  The thermalization status of a heavy quarkonium
system can be inferred from the occupation number distribution of
heavy quarkonium single-particle states.  The occupation numbers in a
thermalized heavy-quarkonium system will obey a Bose-Einstein
distribution characterized by the temperature.  In our present work,
we shall study both a thermalized heavy quarkonium system and an
isolated $Q\bar Q$ bound state in the quark-gluon plasma.

\section{Lattice gauge calculations}

In a quark-gluon plasma, a quarkonium is actually a heavy quark and a
heavy-antiquark each surrounded by a cloud of gluons and quarks.  In
the quenched approximation in which there are no dynamical quarks, the
cloud surrounding the heavy quark and antiquark is approximated to
consist of gluons only.  As gluons are involved, the quark-antiquark
system will be in different color states at different instances.  We
shall be interested in those systems in which the heavy quark and
antiquark exist in the color-singlet state.  Only in the color-singlet
state will be the effective interaction between a quark (plus its
cloud) and an antiquark (plus its cloud) be attractive.  Such a
color-singlet system can further absorb a gluon and become a
color-octet system and we shall also study the cross section for such
a process.

The interaction between a heavy quark and a heavy antiquark in the
color-singlet state was studied by Kaczmarek, Karsch, Petreczky, and
Zantow \cite{Kac03}.  They calculated $\langle tr L({\bf r}/2)
L^\dagger(-{\bf r}/2)\rangle$ in the quenched approximation and they
obtained the color-singlet free energy $F_1({\bf r},T)$ from
\begin{eqnarray}
\label{free1}
\langle tr L({\bf r}/2) L^\dagger(-{\bf r}/2)\rangle = e^{-F_1({\bf
r},T)/kT}.
\end{eqnarray}
Here $tr L( {\bf r}/2) L^\dagger(-{\bf r}/2)$ is the trace of the
product of two Polyakov lines at ${\bf r}/2$ and $-{\bf r}/2$.  The
quark and the antiquark lines do not, in general, form a close loop.
As a gauge transformation introduces phase factors at the beginning
and the end of an open Polyakov line, $\langle tr L( {\bf r}/2)
L^\dagger(-{\bf r}/2)\rangle$ is not gauge invariant under a gauge
transformation.  Calculations have been carried out in the Coulomb
gauge which is the proper gauge to study bound states.

It should be noted that while the interaction between the quark and
the antiquark is gauge dependent, the bound state energies are
physical quantities and they do not depend on the gauge.  As we
explain below, a judicial choice of the Coulomb gauge in the bound
state calculation will help in avoiding spurious next-to-leading
contributions and singularities, which can be removed in other gauges
only by additional laborious effort \cite{App77,Lov78,Bar68}.

To study the bound states of a heavy quarkonium, we need a bound-state
equation, such as the Bethe-Salpeter equation, and the interaction
kernel in the equation.  The non-relativistic approximation of the
Bethe-Salpeter equation leads to the usual Schr\" odinger equation
with the gauge-boson-exchange interaction \cite{Ber82,Oh02}.  It is
necessary to choose a gauge to specify the gauge-boson-exchange
interaction.  We can consider the case of QED from which we can get a
good insight on the gauge dependence.  For the static non-relativistic
problem, the natural choice in the gauge-boson-exchange potential is
the Coulomb gauge, in which the $1/{\bf q}^2$ behavior is found in
single Coulomb photon exchange.  The binding energy, which is of order
$\alpha^2$, has corrections only in the $\alpha^4$ order.  It gives
the correct Breit equation with the proper spin properties when we
expand the interaction to the next order.  Graphs with the cross
two-Coulomb-exchange diagrams vanish in the static limit, and
uncrossed multiple Coulomb exchanges are strictly iterations of the
potential \cite{App77}. In any other gauge, the zero-zero component of
the photon propagator has some residual non-instantaneous
contributions.  A large number of Bethe-Salpeter kernels need to be
included to eliminate the spurious contributions in the next-order of
$\alpha^3$ and $\alpha^3/\ln \alpha$ corrections \cite{Bar68,Lov78}.
Therefore, in their work on the static potential in QCD, Appelquist,
Dine, and Muzinich suggested that the gauge freedom can be used to
eliminate spurious long-range forces at the outset.  They found that
the Coulomb gauge continues to be useful in the static potential in
QCD. The dynamics is now considerably complicated but spurious
contributions are still eliminated \cite{App77}.

Based on the above, it is therefore important to recognize, as in QED,
that there is a gauge-dependence in the two-body bound-state potential
in the Bethe-Salpeter equation but it is most appropriate to solve the
bound state problem using two-body potentials obtained in the Coulomb
gauge, as was obtained by Kaczmarek $et~al$ \cite{Kac03}.

\section{Heavy Quarkonium States in a Thermalized Quark-Gluon Plasma} 

The state of a heavy quarkonium in a quark-gluon plasma can be
described by a density matrix. The set of single-particle states for
this density matrix can be chosen such that they can be represented
well by quarkonium states in a $Q$-$\bar Q$ potential, and the
residual interaction between the gluons with the quarkonium can be
treated as a perturbation.  In this single-particle basis, the heavy
quarkonium density matrix can be approximated to contain only diagonal
matrix elements representing the probabilities for the occupation of
different single-particle states.  What is the $Q$-$\bar Q$ potential
that enters into the Schr\" odinger equation for these quarkonium
single-particle states?

The $Q$-$\bar Q$ potential in perturbative QCD has been studied by
Petreczky \cite{Pet05}.  We would like to examine here the $Q$-$\bar
Q$ potential in non-perturbative lattice QCD calculations.  In
previous analysis, the $Q$-$\bar Q$ potential was taken to be the free
energy $F_1$ for a pair of correlated Polyakov lines, assuming that
the effects of entropy are small \cite{Dig01a,Dig01b,Won01a,Won02}.
It was, however, pointed out by Zantow, Kaczmarek, Karsch and
Petreczky \cite{Kac02,Kar03,Zan03} that the effects of entropy are
large and the total internal energy $U_{1}$, may be used as the
$Q$-$\bar Q$ potential. As the theoretical basis for these suggestions
has not been fully discussed in the literature, we shall go into
details on the proper description of the $Q$-$\bar Q$ potential and
single-particle states.

We start first by studying the auxiliary problem of a {\it static}
color-singlet $Q$-$\bar Q$ pair at a separation ${\bf r}$ in a
thermalized quark-gluon plasma.  In the quenched approximation, the
color-singlet free energy $F_1({\bf r},T)$ for such a static pair, can
be written from Eq.\ (\ref{free1}) explicitly in the Euclidean time
$\tau=it$ as \cite{Yaf82}

\begin{eqnarray}
\label{neq2}
e^{-\beta F_1({\bf r},T)}= 
Z({\bf r},T) \biggr / Z_0(T),
\end{eqnarray}
\begin{eqnarray}
Z({\bf r},T)
=\int [dA] W_{Q\bar Q}(A,T, {\bf r}), 
\end{eqnarray}
\begin{eqnarray}
Z_0(T)=\int [dA] W_0(A,T),
\end{eqnarray}
\begin{eqnarray}
 W_{Q\bar Q}(A,T,{\bf r}) = tr\left \{ \hat P \exp\{ \int_0^\beta
g(\tau) d \tau A_0( \frac {{\bf r}}{2}, \tau)\}  
\exp\{\int_\beta^0 g(\tau) d \tau A_0( -\frac{{\bf r}}{2}, \tau )\} \right
\} W_0(A,T),
\end{eqnarray}
\begin{eqnarray}
W_0(A,T) = \exp\left \{-\frac {1}{4} \int_0^\beta d\tau \int d^3 x
F_{\mu \nu} F^{\mu \nu} \right \},
\end{eqnarray}
where $\beta=1/kT$ is the inverse temperature, $Z({\bf r},T)$ is the
partition function when a color-singlet $Q$ and $\bar Q$ separated by
a distance ${\bf r}$ is placed in the gluon medium, $Z_0(T)$ is the
partition function in the absence of $Q$ and $\bar Q$.  The operator
${\hat P}$ is the path-order operator, which is the time-order
operator ${\hat T}$ for $\exp\{\int_0^\beta g(\tau) d \tau A_0( {{\bf
r}}/{2}, \tau )\}$ and is the reverse time-order operator for
$\exp\{\int_\beta^0 g(\tau) d \tau A_0(-{{\bf r}}/{2}, \tau )\}$.  The
free energy $F_1({\bf r},T)$ with the $Q$-$\bar Q$ pair is measured
relative to the free energy $F_{0}(T)$ without the $Q$-$\bar Q$ pair.
We rewrite Eq.\ (\ref{neq2}) as
\begin{eqnarray}
\label{ef}
Z_0(T) =\int [dA]  
e^{\beta F_1({\bf r},T)}
tr\left \{ 
\hat P \exp\{ \int_0^\beta g(\tau) d \tau A_0(\frac {{\bf r}}{2}, \tau)\}
\exp\{\int_\beta^0 g(\tau) d \tau A_0(-\frac{{\bf r}}{2}, \tau)\} 
\right \}
\exp \left \{
-\frac {1}{4} \int_0^\beta  d\tau \int d^3 x F_{\mu \nu} F^{\mu \nu} 
\right \}.
\end{eqnarray}
Taking the derivative of this equation with respect to $\beta$, we
obtain for the derivative of the left-hand side
\begin{eqnarray}
\label{neq3}
\frac {\partial \{{\rm LHS}\}}{\partial \beta} 
=\frac{\partial Z_0(T)}{\partial \beta}
=\int [dA]  \left \{ - \frac {1}{4} \int d^3 x  F_{\mu \nu} F^{\mu \nu}  \right \} 
W_{0}(A,T),
\end{eqnarray}
and for the derivative of the right-hand side, we get
\begin{eqnarray}
\label{neq3a}
& &\frac {\partial \{{\rm RHS}\}}{\partial \beta} 
= \int [dA]  e^{\beta F_1({\bf r},T)}
\Biggl [ \left \{ F_1({\bf r},T) +\beta \frac {\partial  F_1({\bf r},T)}{\partial \beta}
- \frac {1}{4} \int d^3 x  F_{\mu \nu} F^{\mu \nu}  \right \} 
W_{Q\bar Q}(A,T,{\bf r})
\nonumber\\
& & + 
tr \left \{ g(T) (  A_0(\frac {{\bf r}}{2}, T)
-  A_0(-\frac{{\bf r}}{2}, T)) 
 {\hat P} \exp\{ \int_0^\beta g(\tau) d \tau A_0(\frac {{\bf r}}{2}, \tau)\}
 \exp\{\int_\beta^0 g(\tau) d \tau A_0(-\frac{{\bf r}}{2}, \tau)\} 
\right \}
W_{0}(A,T) \Biggr ].
\end{eqnarray}
We equate the above Eq.\ (\ref{neq3}) to Eq.\ (\ref{neq3a}). Using
$e^{\beta F_1({\bf r},T)}=Z_0(T)/Z({\bf r},T)$ and dividing the
resultant equation by $Z_0(T)$, we obtain then the proper
thermodynamic equality relating $F_1$, $S_1$, and $U_1$, for a system
with a color-singlet $Q$ and $\bar Q$ separated ${\bf r}$ at
temperature $T$,
\begin{eqnarray}
\label{FF}
F_1({\bf r},T)+TS_1({\bf r},T)=U_{1} ({\bf r},T),
\end{eqnarray}
where $S_1({\bf r},T)=-\partial F_1({\bf r},T)/\partial T$ is the
color-singlet entropy with the $Q$-$\bar Q$ pair and is measured
relative to the entropy $S_0(T) =-\partial F_0(T)/\partial T$ without
the $Q$-$\bar Q$ pair, and $U_1({\bf r},T)$ is the total color-singlet
internal energy given explicitly by
\begin{eqnarray}
\label{Uqq}
U_{1} ({\bf r},T)=U^{(1)}_{Q\bar Q}({\bf r},T)+  U^{(1)}_g({\bf r},T)
- U_{g0}(T),
\end{eqnarray}
\begin{eqnarray}
U_{g0}(T)
=
\int [dA] 
\left \{\frac {1}{4} \int d^3 x F_{\mu \nu} F^{\mu \nu} \right \}
W_0(A,T) \Biggr / \int [dA] W_0 (A,T),
\end{eqnarray}
\begin{eqnarray}
U^{(1)}_g ({\bf r},T)
=\int [dA] \left \{\frac {1}{4} \int d^3 x F_{\mu \nu} F^{\mu \nu} \right \}
W_{Q\bar Q} (A,T,{\bf r}) \Biggr / \int [dA] W_{Q\bar Q} (A,T,{\bf r}),
\end{eqnarray}
and
\begin{eqnarray}
\label{lat2}
U^{(1)}_{Q\bar Q}({\bf r},T)
&=&   
\int [dA]  
tr \left \{ g(T)  [A_0(\frac {{\bf r}}{2}, T)
-   A_0(-\frac{{\bf r}}{2}, T)] 
\hat P  \exp\{ \int_0^\beta g(\tau) d \tau A_0(\frac {{\bf r}}{2}, \tau)\}
 \exp\{\int_\beta^0 g(\tau) d \tau A_0(-\frac{{\bf r}}{2}, \tau)\} 
\right \} W_{0}(A,T)
\nonumber\\
&\div&
\int [dA]  
tr \left \{ \hat P
 \exp\{ \int_0^\beta g(\tau) d \tau A_0(\frac {{\bf r}}{2}, \tau)\}
 \exp\{ \int_\beta^0 g(\tau) d \tau A_0(-\frac{{\bf r}}{2}, \tau)\} 
\right \} W_{0}(A,T),
\end{eqnarray}

We may attempt to give names to various mathematical expressions.  In
Euclidean time, the quantity $ F_{\mu \nu} F^{\mu \nu}/4$ is equal to
$({\bf E}^2+{\bf B}^2)/2$, the gluon energy density
\cite{Yaf82,Fey65}.  The quantity $U_{g0}(T)$ is the expectation value
of $\int d^3x ({\bf E}^2+{\bf B}^2)/2$ with the weight function
$W_0(A,T)$, and it corresponds to the gluon internal energy in the
absence of the heavy quark pair.  It is independent of the separation
${\bf r}$ between $Q$ and $\bar Q$.  In contrast, $U_g^{(1)}({\bf
r},T)$ is the expectation value of $\int d^3x ({\bf E}^2+{\bf B}^2)/2$
with the weight function $W_{Q\bar Q}(A,T,{\bf r})$, and it
corresponds to the gluon internal energy in the presence of the heavy
quark pair.  Consequently, $U_g^{(1)}({\bf r},T)$ depends on the
separation ${\bf r}$ between $Q$ and $\bar Q$.  The difference between
the total internal energy $U_1$ and gluon internal energy difference
$U_g^{(1)}({\bf r},T)-U_{g0}(T)$ is the quantity $U^{(1)}_{Q\bar
Q}({\bf r},T)$, the internal energy of the heavy-quark pair, including
the interaction between $Q$ and $\bar Q$ as well as the interaction
between $Q$ with gluons and $\bar Q$ with gluons.

Eqs.\ (\ref{FF})-(\ref{lat2}) show that the total internal energy
$U^{(1)}({\bf r},T)$ contains the gluon internal energy difference
$U_g^{(1)}({\bf r},T)- U_{g0}(T)$. In order to obtain the ${\bf
r}$-dependence of the internal energy of the heavy quark pair
$U_{Q\bar Q}^{(1)}({\bf r},T)$, it is necessary to subtract the gluon
internal energy difference $U_g^{(1)}({\bf r},T)-U_{g0}(T)$ from the
total internal energy $U_1({\bf r},T)$. In Section VI we shall show a
method to carry out such a subtraction in the local energy-density
approximation.
 
\section{Equation of motion for $Q$-$\bar Q$ single particle states}

Lattice gauge calculations provide information on the free energy
$F_1$ and the total internal energy $U_1$ for a static color-singlet
$Q$ and $\bar Q$ separated by a distance ${\bf r}$.  Quantities for
$Q$ and $\bar Q$ in the color-octet state can be similarly obtained.
For simplicity, we shall limit our consideration to a system of
color-singlet $Q \bar Q$ states. The generalization to a system
color-octet states can be easily carried out.

We would like to use an appropriate variational principle to obtain
the equation of motion for the color-singlet quarkonium states in a
quark-gluon plasma.  For such a purpose, we consider a schematic toy
model that retains the relevant features of the system.  In the
quenched approximation without dynamical light quarks, the quark-gluon
plasma consists of gluons only.  The quark $Q$ and antiquark $\bar Q$
are in dynamical motion in different single-particle states.  For
simplicity, we presume that the color degree of freedom has been
integrated out. The dynamical degrees of freedom in our schematic
model are then the $Q\bar Q$ and gluon states, $\psi_i({\bf r})$ and
$\phi_i({\bf r})$, which can be bound or unbound, the corresponding
$Q\bar Q$ and gluon state occupation numbers, $n_i(Q\bar Q)$ and
$n_j(g)$, and the total number of gluons $N_g$.  In the schematic toy
model, we represent the $Q$-$\bar Q$, $Q$-$g$, $\bar Q$-$g$, and
$g$-$g$ interactions when $Q$ and $\bar Q$ belong to a color-singlet
state by $V_{Q\bar Q}$, $V_{Q g}$, $V_{\bar Q g}$, and $V_{gg}$
respectively.

The equilibrium of the system at a constant temperature $T$ is
characterized by minimizing the grand potential $A$ appropriate for
the color-singlet $Q$ and $\bar Q$ in dynamical motion in the
quark-gluon plasma given by
\begin{eqnarray}
\label{eqA}
A={\cal F}_1-\mu(Q\bar Q) N_{Q\bar Q} - \mu(g)N_g ={\cal U}_{1} - T{\cal S}_1
-\mu(Q\bar Q) N_{Q\bar Q} - \mu(g)N_g,
\end{eqnarray}
where ${\cal F}_1$ is free energy, ${\cal U}_1$ the total internal
energy, and ${\cal S}_1$ the total entropy for $Q$ and $\bar Q$ in
dynamical motion in the quark-gluon plasma.  Only in their static
limits when the heavy quark and antiquark are held spatially fixed can
${\cal F}_1$, ${\cal U}_1$ and ${\cal S}_1$ be equal to their
corresponding static thermodynamical quantities $F_1$, $U_1,$ and
$S_1$.  The quantities $\mu(Q\bar Q)$ and $\mu(g)$ are the chemical
potentials of $Q\bar Q$ and gluons respectively, and $N_{Q\bar Q}$ and
$N_g$ are the numbers of $Q\bar Q$ and gluons respectively.  Strictly
speaking the number of gluons at thermal equilibrium depends on the
length scale (the $Q^2$ value of the measuring probe) under
consideration.  In our schematic model, we can fix the length scale
appropriate for $Q$-$\bar Q$ bound states in the quark-gluon plasma,
and in that length scale the number of gluons, for the fixed spatial
volume under consideration at thermal equilibrium at $T$, can be
determined.  We shall also ignore the annihilation of $Q$-$\bar Q$
into light quarks or photons and the corresponding inverse production
so that the number of $Q\bar Q$ pair can be considered fixed also.  In
the above equation, we need to add the Lagrange multipliers
$\lambda_i(Q\bar Q)$ and $\lambda_j(g)$ to the grand potential in
order to constrain the normalization of the wave functions.

To carry out the minimization of the grand potential to obtain the
equation of motion of the single-particle states, we can follow
Bonche, Levit, and Vautherin \cite{Clo68,Bal80,Bon81,Bon84,Bon85} and
write down the grand potential $A$ explicitly.  The internal energy
${\cal U}_1$ in Eq.\ (\ref{eqA}) is ${\cal U}_{Q\bar Q}^{(1)}+{\cal
U}_g^{(1)}-U_{g0}$, the sum of the internal energy of the heavy quark
pair and the gluons, relative to the gluon internal energy when the
heavy quark pair is not present.  In terms of the wave function and
occupation number degrees of freedom, the grand potential $A$ can be
written explicitly as
\begin{eqnarray}
A&=&
\sum_i n_i(Q\bar Q) \int d{\bf r} \, \psi_i^{\dagger}({\bf r})
[ \frac{\hbar^2 {\bf p}^2} {2 \mu_{\rm red}} 
+V_{Q \bar Q} ({\bf r})] \psi_i({\bf r})
+\sum_{i,j} n_i(Q\bar Q) n_j(g)
\langle \psi_i \phi_j |V_{Qg}+V_{\bar Q g}|\psi_i \phi_j\rangle
\nonumber\\
&+& \sum_j n_j(g) \int d{\bf r}' \phi_j^{\dagger}({\bf r}')
\sqrt{ {\bf p}_g^2 +m_{\rm eff}^2}~ \phi_j({\bf r}') 
 + \sum_{j,k} n_j(g)       n_k(g)\langle \phi_j \phi_k |V_{gg}|
    \phi_{j} \phi_{k}+ \phi_{k} \phi_{j}\rangle/2  - U_{g0}
\nonumber\\
&+& T\sum_i [ n_i(Q\bar Q) \ln  n_i(Q\bar Q)
-\left \{ 1+ n_i(Q\bar Q) \right \} \ln  \left \{ 1+ n_i(Q\bar Q) \right \}]
\nonumber\\
&+&T\sum_j [ n_j(g) \ln  n_j(g)
- \left \{ 1+n_j(g) \right \} \ln \left \{ 1+ n_j(g)\right \}]
\nonumber\\
&-& \mu(Q\bar Q) \sum_i n_i(Q \bar Q)  - \mu (g) \sum_j n_j(g) 
- \sum_i \lambda_i(Q\bar Q) \langle \psi_i | \psi_i\rangle
- \sum_j \lambda_j(g)       \langle \phi_j | \phi_j\rangle.
\end{eqnarray}
Here ${\bf p}$ and $\mu_{\rm red}$ are the relative momentum and the
reduced mass of $Q$-$\bar Q$, ${\bf p}_g$ and $m_{\rm eff}$ are the
momentum and the effective mass of the gluon, and the dependence of
various quantities on the temperature is made implicit.  In this
expression for the grand potential, the first two terms give ${\cal
U}_{Q\bar Q}^{(1)}$, the third and fourth terms give ${\cal
U}_{g}^{(1)}$, and the sixth and seventh terms give the entropy
$T{\cal S}_1$. The matrix element $\langle \psi_i \phi_j
|V_{Qg}+V_{\bar Q g}|\psi_i \phi_j\rangle$ is
\begin{eqnarray}
\langle \psi_i \phi_j |V_{Qg}+V_{\bar Q g}|\psi_i \phi_j\rangle 
=\int d{\bf r}\, d{\bf r}' \, 
 \psi_i^{\dagger}({\bf r}) \phi_j^{\dagger}({\bf r}')
[V_{Q g} ({\bf r}'+\frac{\bf r}{2})+V_{\bar Q g} ({\bf r}'-\frac{\bf r}{2})]
\psi_i({\bf r}) \phi_j({\bf r}'),
\end{eqnarray}
and the matrix element $\langle \phi_j \phi_k |V_{gg}| \phi_{j}
\phi_{k}+ \phi_{k} \phi_{j}\rangle$ is
\begin{eqnarray}
\langle \phi_j \phi_k |V_{gg}|\phi_j \phi_k +\phi_j \phi_k \rangle 
=\int d{\bf r}'\, d{\bf r}'' \, 
\phi_j^{\dagger}({\bf r}')\phi_k^{\dagger}({\bf r}'')
V_{g g} ({\bf r}'-{\bf r}'')
[\phi_j({\bf r}') \phi_k({\bf r}'')+\phi_k({\bf r}') \phi_j({\bf r}'')].
\end{eqnarray}
When we carry out the minimization of the grand potential with respect
to the dynamical degrees of freedom, we obtain five equations (Eqs.\
(24)-(28) below).  By minimizing with respect to $\psi^\dagger_i ({\bf
r})$, we obtain the equation of motion for $\psi_i({\bf r})$ of the
$Q\bar Q$ system as
\begin{eqnarray}
\label{sch}
\left \{ \frac{\hbar^2 {\bf p}^2} {2 \mu_{\rm red}} +V_1({\bf r},T) 
-\epsilon_i' (Q\bar Q) \right \} \psi_i({\bf r}) = 0,
\end{eqnarray}
where the color-singlet single-particle potential $V_1({\bf r},T)$ is
\begin{eqnarray}
V_1({\bf r},T) = V_{Q \bar Q}
({\bf r}) + \sum_j n_j(g) \int d{\bf r}' \phi_j^{\dagger}({\bf r}')
[V_{Q g} ({\bf r}'+{\bf r}/2)+V_{\bar Q g} ({\bf r}'-{\bf r}/2)],
\end{eqnarray}
and $\epsilon_i'(Q\bar Q)=\lambda_i(Q \bar Q)/n_i(Q \bar Q)$.  We note
that in the above equation, the quantity
\begin{eqnarray}
\label{eq1a}
\sum_j n_j (g) \phi_j^{\dagger} ({\bf r}')
 \phi_j ({\bf r}')
=\rho_g ({\bf r}')
\end{eqnarray}
is the gluon density $\rho_g ({\bf r}')$.  The color-singlet potential
$V_1 ({\bf r},T)$ is
\begin{eqnarray}
V_1 ({\bf r},T)= 
V_{Q\bar Q}({\bf r},T)+\int 
d {\bf r}' \rho_g({\bf r}',T) 
[V_{Q g}({\bf r}'+\frac{\bf r}{2})
+V_{{\bar Q} g}({\bf r}'-\frac{\bf r}{2})],
\end{eqnarray}
which represents the internal energy of a static color-singlet
$Q$ and $\bar Q$ at a separation ${\bf r}$ in the quark-gluon plasma.
It has the same physical meaning as internal energy $U_{Q\bar
Q}^{(1)}({\bf r},T)$ of a static color-singlet $Q$ and $\bar Q$ at a
separation ${\bf r}$ in the lattice gauge calculations.  It is
therefore reasonable to identify the $Q\bar Q$ internal energy
$U_{Q\bar Q}^{(1)}({\bf r},T)$ of the lattice gauge calculations as
the single-particle potential $V_1({\bf r},T)$ in the Schr\" odinger
equation (\ref{sch}), and write it as
\begin{eqnarray}
\label{eq1}
\left \{ \frac{\hbar^2 {\bf p}^2} {2 \mu_{\rm red}} + U_{Q\bar Q}^{(1)}(
{\bf r},T) - \epsilon_i' (Q\bar Q) \right \} \psi_i({\bf r}) = 0.
\end{eqnarray}
It is simplest to re-calibrate the energy (and similarly, the chemical
potential) by $\epsilon_i(Q\bar Q) =\epsilon_i'(Q\bar Q)-U_{Q\bar
Q}({\bf r})_{r \to \infty}$ and re-write the above equation as
\begin{eqnarray}
\label{exact}
\left \{ \frac{\hbar^2 {\bf p}^2} {2 \mu_{\rm red}} + 
U_{Q\bar Q}^{(1)}({\bf r},T) -  U_{Q\bar Q}^{(1)}(|{\bf r}|\to \infty,T) - 
\epsilon_i (Q\bar Q) \right \} \psi_i({\bf r}) = 0.
\end{eqnarray}
The above Schr\" odinger equation involving $U_{Q\bar Q}^{(1)}({\bf
r},T)$ is the proper equation of motion for quarkonium single-particle
states.

From the above considerations, the $Q$-$\bar Q$ potential in the
equation of motion of quarkonium single-particle states is $U_{Q\bar
Q}^{(1)}({\bf r},T)$.  Lattice calculations so far provide information
only on the free energy $F_1({\bf r},T)$ and the total internal energy
$U_{1}({\bf r},T)$.  It will be of great interest in future lattice
work to evaluate $U_{Q\bar Q}^{(1)}({\bf r},T)$ so that it can be used
as the proper $Q$-$\bar Q$ potential in quarkonium studies. In the
next section, we shall present a method by which $U_{Q\bar
Q}^{(1)}({\bf r},T)$ can be approximately evaluated by using the
quark-gluon plasma equation of state.

The minimization of the grand potential with respect to the gluon wave
function $\phi^{\dagger}_j({\bf r})$ gives the equation of motion for
the gluon states,
\begin{eqnarray}
\label{eq2}
& &\Biggl \{\sqrt{{\bf p}_g^2 + m_{\rm eff}^2}
+\sum_i n_i(Q\bar Q) \int  d{\bf r} \psi_i^{\dagger}({\bf r})
[V_{Q g} ({\bf r}'+{\bf r}/2)+V_{\bar Q g} ({\bf r}'-{\bf r}/2)]
\psi_i({\bf r})
-\epsilon_i(g)  \Biggr \}
\phi_j({\bf r}')  
\nonumber\\
&+&\sum_{k} n_k(g) \int d{\bf r}'' \phi_k^\dagger ({\bf r}'')
V_{gg}({\bf r}''-{\bf r}')
[ \phi_{k} ({\bf r}'') \phi_j ({\bf r}') +\phi_{k} ({\bf r}') 
\phi_j ({\bf r}'')]
\nonumber\\
&+&
\sum_{\lambda,k} n_\lambda(g)  n_k(g) 
\int d{\bf r}''
 \phi_{\lambda}^\dagger ({\bf r}')  \phi_{k}^\dagger ({\bf r}'') 
\frac{\partial V_{gg}({\bf r}''-{\bf r}')}{\partial \rho_g}
[ \phi_{k} ({\bf r}'') \phi_\lambda ({\bf r}') +\phi_{k} ({\bf r}') 
\phi_\lambda ({\bf r}'')]
n_j(g)\phi_j({\bf r}')
= 0,
\end{eqnarray}
where we have taken the density-dependent interaction as a
delta-function in ${\bf r}''-{\bf r}'$ as in the Skyrme interaction
\cite{Vau72,Won79}.  The minimization of the grand potential with
respect to $n_i(Q\bar Q)$ and $n_j(g)$ yields
\begin{eqnarray}
\label{be1}
n_i(Q\bar Q)=\frac {1} {e^{[\epsilon_i (Q\bar Q)-\mu(Q\bar Q)]/T}-1},
\end{eqnarray}
and
\begin{eqnarray}
\label{be2}
n_j(g)=\frac {1} {e^{[\epsilon_j (g)      -\mu(g)]/T}      -1},
\end{eqnarray}
which are the well-known Bose-Einstein distributions.

The variation of the grand potential $A$ of Eq. (15) with respect to
$N_g$ gives
$$
\partial A/\partial N_g=\mu(g).
$$
The requirement that thermal equilibrium is reached when the variation
of $A$ with respect to $N_g$ is a minimum leads to \cite{Lan80}
\begin{eqnarray}
\mu(g)=0.
\end{eqnarray}
The
number of gluons at level $j$, $n_j(g)$, can then be obtained from
Eq.\ (27) with $\mu(g)=0$, and the total number of gluons $N_g$ is given by
\begin{eqnarray}
N_g=\sum_j n_j(g).
\end{eqnarray}

The above considerations give the set of equations (\ref{exact}),
(\ref{eq2}), (\ref{be1}), (\ref{be2}), and (28) for a system in which
both the quark-gluon plasma and the $Q\bar Q$ are in thermal
equilibrium.  If a quarkonium is placed in a thermalized quark-gluon
plasma for a period longer than the time needed for it to thermalize,
the heavy quarkonium will reach thermal equilibrium and the $Q\bar Q$
system will be described by the above set of single-particle states
with the Bose-Einstein distribution of occupation numbers.

Other cases of our interest are those in which the quark-gluon plasma
has reached thermal equilibrium while the $Q\bar Q$ system, which
arises from the independent mechanism of nucleon-nucleon hard
scattering, may not have.  For such a case, the equations of motion
for the gluons states and gluon occupation numbers, Eqs. (\ref{eq2})
and (\ref{be2}), remain valid, except that in Eq.\ (\ref{eq2}) the
$Q\bar Q$ occupation numbers $n_i(Q\bar Q)$ will no longer obey the
Bose-Einstein distribution (\ref{be2}).  The equation of motion for
the $Q\bar Q$ single-particle states, Eq.\ (\ref{exact}), remains
valid, as it depends on the gluon density. Because the heavy-quark
pair is a rare occurrence and gluons are much greater in number, we
expect that the gluon density and wave functions obtained in
Eq. (\ref{eq2}) is insensitive to the status of $Q\bar Q$
thermalization, and they depend mainly on the thermalization of the
gluons themselves.  Therefore, single-particle states of the $Q\bar Q$
system, $\psi_i({\bf r})$, which depend on the gluon density as given
by Eq.\ (\ref{eq1}), are relatively insensitive to the thermalization
status of $Q\bar Q$, and they depend mainly on the thermalization
status of the gluons.

In view of the above, the set of $Q\bar Q$ single-particles states
$\psi_i({\bf r})$ of Eq.\ (\ref{exact}) can be used to examine the
states of a $Q\bar Q$ system, whether the $Q\bar Q$ has reached
thermal equilibrium or not.  For example, one can introduce a bound
$Q\bar Q$ state $\psi_\lambda$ into a thermalized quark-gluon plasma
of temperature $T$.  Such a system is described by a density matrix
with an initial occupation number $n_i(Q\bar Q)=\delta_{i\lambda}$ at
$t=0$. The collision of the $Q\bar Q$ with gluons in the medium will
lead to the evolution of the occupation numbers of the $Q\bar Q$
states as a function of time, leading eventually to the Bose-Einstein
distribution characterized by the temperature of the medium.

The $Q\bar Q$ system can exist in color-singlet and color-octet
states.  The consideration we have given can be generalized to
color-octet states.  The Schr\" odinger equation for the different
single-particle color states $a$ will depend on $U_{Q\bar
Q}^{(a)}({\bf r},T)$.  For the same reason that the $Q\bar Q$
single-particle states depend on the thermalization status of the
gluons and are insensitive to the thermalization status of $Q\bar Q$,
these single-particle states can be used to examine a $Q\bar Q$ system
in thermalized gluon matter. When the $Q\bar Q$ also reaches thermal
equilibrium, the set of states with occupation numbers given by
equations (\ref{be1}) will include both color-singlet and color-octet
states.

\section{Color-singlet $Q$-$\bar Q$ potential }

From the variational principle of minimizing the grand potential, we
find that the Schr\" odinger equation (\ref{exact}) contains the
$Q$-$\bar Q$ internal energy $U_{Q\bar Q}^{(1)}$ as the quarkonium
$Q$-$\bar Q$ potential.  Lattice calculations so far provide
information only on the free energy $F_1({\bf r},T)$ and the total
internal energy $U_{1}({\bf r})$ but not $U_{Q\bar Q}^{(1)}$.  While
we await accurate lattice gauge results for $U_{Q\bar Q}^{(1)}$, we
can obtain an approximate $U_{Q\bar Q}^{(1)}$ from $F_1$ and $U_1$ by
using the equation of state of the quark-gluon plasma and the First
Law of Thermodynamics.

In the quenched approximation, the quark-gluon plasma consists of
gluons only.  We shall use the following strategy to obtain $U_{Q\bar
Q}^{(1)}$.  From lattice gauge calculations, we can calculate
$F_1-U_1$ which is equal to the entropy content $TS_1$ of the whole
system.  As the heavy quark $Q$ and antiquark $\bar Q$ are held fixed
in the lattice calculation, the entropy $TS_1$ comes entirely from the
deconfined gluons, and it contains no contribution from the heavy
quark $Q\bar Q$ pair.  From the entropy content of the gluon medium
$TS_1$, we can deduce approximately the gluon internal energy
$U_g^{(1)}$ by using the gluon medium equation of state and the First
Law of Thermodynamics.  Then, the knowledge of the approximate gluon
internal energy $U_g^{(1)}$ and the total internal energy $U_1$ gives
a color-singlet potential $U_{Q \bar Q}^{(1)}$ in terms of $F_1$ and
$U_1$.

Following such a strategy, we consider a deconfined gluon medium at
temperature $T$ with a static color-singlet $Q$ and $\bar Q$ separated
by a distance ${\bf r}$.  We focus our attention on a volume element
$dV$ of gluons at spatial position ${\bf x}$ in which the gluon
internal energy element is $dU_g^{(1)}$ and the gluon entropy element
is $dS_1$.  At this spatial location, a gluon experiences an external
potential ${\cal V}({\bf x})$ exerted by the color-singlet $Q$ and
$\bar Q$, other constituents, and induced charges.  Hydrostatic
equilibrium is reached when the local gluon pressure $p^{(1)}({\bf
x})$ at the volume element counterbalances the external forces.  We
can begin in the non-relativistic description in which the total external
force acting on a unit volume at ${\bf x}$ is $\rho_g({\bf x})
\nabla_{\bf x}{\cal V}({\bf x})$ and hydrostatic equilibrium is
reached when \cite{Won75}
\begin{eqnarray}
\nabla_{\bf x}p^{(1)}({\bf x}) + \rho_g({\bf x}) \nabla_{\bf x}{\cal V}({\bf x})=0.
\end{eqnarray} 
In the relativistic case, one can describe the external interaction
${\cal V}({\bf x})$ as part of the $g^{00}$ metric,
$g^{00}=1+2\rho_g({\bf x}){\cal V}({\bf x})/\epsilon({\bf x})$, and
hydrostatic equilibrium is reached when
\cite{Whe70,Wei72}
\begin{eqnarray}
\nabla_{\bf x}p^{(1)}({\bf x}) +\{p^{(1)}({\bf x})+\epsilon_g^{(1)}({\bf x})\}
 \nabla_{\bf x}\sqrt{1+2\rho_g({\bf x}){\cal V}({\bf x})/\epsilon_g^{(1)}({\bf x})}=0.
\end{eqnarray} 
Thus, the external forces determine the gluon pressure $p^{(1)}({\bf
x})$ that is required to maintain hydrostatic equilibrium.  At
thermal equilibrium, the local gluon pressure is related to the local
gluon entropy and the local gluon energy density. The external force
therefore leads to a change of the gluon local entropy and the local
gluon energy density.

According to this picture, the change in entropy is zero when the
heavy quark sits on top of the heavy antiquark at ${\bf r}=0$, as the
color charges will neutralize for a color-singlet $Q\bar Q$.  As $Q$
separates from $\bar Q$, the local gluon pressure, energy density, and
entropy will increase in the vicinity of $Q$ and $\bar Q$ to
counterbalance the forces due to $Q$ and $\bar Q$.  The increase in
entropy will reach a constant value at large separations when the $Q$
and $\bar Q$ each independently causes a modification of the gluon
distribution in its vicinity.

Under the temperature $T$ and pressure $p^{(1)}({\bf x})$, the
gluon internal energy element $dU_g^{(1)}$ and the gluon entropy
element $dS_1$ at ${\bf x}$ are related by the First Law of
Thermodynamics,
\begin{eqnarray}
dU_g^{(1)}=TdS_1-p^{(1)}({\bf x})dV,
\end{eqnarray}
where the superscript $(1)$ denotes that the heavy quark $Q\bar Q$
pair is in a color-singlet state.  The local gluon internal energy
density $\epsilon_g^{(1)}({\bf x})$ is therefore related to the local
gluon entropy density ${dS_1}/{dV}$ and local gluon pressure
$p^{(1)}({\bf x})$ by
\begin{eqnarray}
\epsilon_g^{(1)}({\bf x})
=  \frac{dU_g^{(1)}}{dV}({\bf x})
=T\frac{dS_1}{dV}({\bf x}) - p^{(1)}({\bf x}).
\end{eqnarray}
The equation of state of a homogeneous quark-gluon plasma (gluon medium
only in the case of the quenched approximation) has been obtained in
previous lattice calculations \cite{Boy96}.  It can be represented by
expressing the ratio $p/(\epsilon/3)$ as a function $a(T)$ of
temperature,
\begin{eqnarray}
a(T)= \frac{p(T)}{\epsilon_g(T)/3}.
\end{eqnarray}
We shall make the local energy-density approximation in which the
local gluon energy density $\epsilon_g^{(1)}({\bf x})$ and the local
gluon pressure $p^{(1)}({\bf x})$ under the temperature $T$ obey the
equation of state for the (homogeneous) bulk quark-gluon plasma so
that
\begin{eqnarray}
\label{local}
\frac{p^{(1)}({\bf x})}
{\epsilon_g^{(1)}({\bf x})/3}=a(T),  
\end{eqnarray}
as in the usual hydrodynamical description of the quark-gluon
plasma. We then have
\begin{eqnarray}
\label{dug}
\frac{dU_g^{(1)}}{dV}({\bf x})
=\frac{3}{3+a(T)} \frac{TdS_1}{dV}({\bf x}).
\end{eqnarray}
When we integrate the local gluon energy density over the whole volume
of ${\bf x}$ under consideration, we obtain
\begin{eqnarray}
\label{lenergy}
U_g^{(1)}({\bf r},T) = \int d{\bf x} 
\frac{dU_g^{(1)}}{dV}({\bf x})
=\frac{3}{3+a(T)} \int d{\bf x} \frac{TdS_1}{dV}({\bf x})
=\frac{3}{3+a(T)} TS_1({\bf r},T).
\end{eqnarray}

In a lattice gauge calculation, the large degrees of freedom of the
gauge field and the maintenance of a thermal bath of constant
temperature $T$ make it reasonable to assume local thermal equilibrium
and use the equation of state of homogeneous bulk matter to relate the
local energy density to the local pressure in the local energy-density
approximation.  It will be of great interest to check the above
equation (\ref{lenergy}) in future lattice gauge calculations in order
to test the validity of the local energy-density approximation.

For the case in the absence of the $Q$-$\bar Q$ pair, the gluon
internal energy $U_{g0}(T)$ and the gluon entropy $S_0(T)$ are
similarly related by
\begin{eqnarray}
U_{g0}(T) =\frac{3}{3+a(T)} TS_0(T).
\end{eqnarray}
We therefore have
\begin{eqnarray}
\label{oldnot}
U_g^{(1)}({\bf r},T) - U_{g0}(T) =\frac{3}{3+a(T)} T\left \{
S_1({\bf r},T)-S_0(T) \right \}.
\end{eqnarray}

In the above discussions from Eq.\ (32) to Eq.\ (39), we have followed
the standard practice to use the convention of ``absolute'' units in
which the entropy $S_1({\rm absolute})$ and $S_0({\rm absolute})$ are
measured relative to zero entropy.  On the other hand, in lattice
calculations and in Section IV, we have used the ``lattice gauge
calculation'' convention in which the entropy $S_1({\rm lattice})$ and
the free energy $F_1({\rm lattice})$ with the $Q$-$\bar Q$ pair are
measured relative respectively to the entropy $S_0({\rm absolute})$
and free energy $F_0({\rm absolute})$ without the $Q$-$\bar Q$ pair
(i.e. $S_1({\rm lattice})=S_1({\rm absolute})-S_0({\rm absolute})$ and
$F_1({\rm lattice})=F_1({\rm absolute})-F_0({\rm absolute})$).
Henceforth, we shall switch back to the ``lattice gauge calculation''
convention in which the entropy $S_1({\bf r},T)$ and the free energy
$F_1({\bf r},T)$ with the the $Q$-$\bar Q$ pair are measured relative
to those in the absence of the $Q$-$\bar Q$ pair.  In this lattice
gauge convention, the above equation (\ref{oldnot}) can be re-written
as
\begin{eqnarray}
U_g^{(1)}({\bf r},T) - U_{g0}(T) =\frac{3}{3+a(T)} TS_1({\bf r},T).
\end{eqnarray}

Using the relation $TS_1=U_1-F_1$ of Eq.\ (10) in lattice gauge
theory, we can express the above $U_g^{(1)}({\bf r},T)-U_{g0}(T)$ in
terms of $F_1({\bf r},T)$ and $U_1({\bf r},T)$,
\begin{eqnarray}
U_g^{(1)}({\bf r},T) - U_{g0}(T)
=\frac{3}{3+a(T)} \{U_1({\bf r},T)-F_1({\bf r},T)\}. 
\end{eqnarray}
Substituting this equation in Eq.\ (\ref{Uqq}), 
we obtain 
\begin{eqnarray}
\label{Uqq1}
U_{Q\bar Q}^{(1)}({\bf r},T)= 
 \frac{3}{3+a(T)}    F_1({\bf r},T) 
+\frac{a(T)}{3+a(T)} U_1({\bf r},T).
\end{eqnarray}

\vspace*{-0.0cm}
\hspace*{1.5cm}
\begin{figure} [h]
\includegraphics[angle=0,scale=0.50]{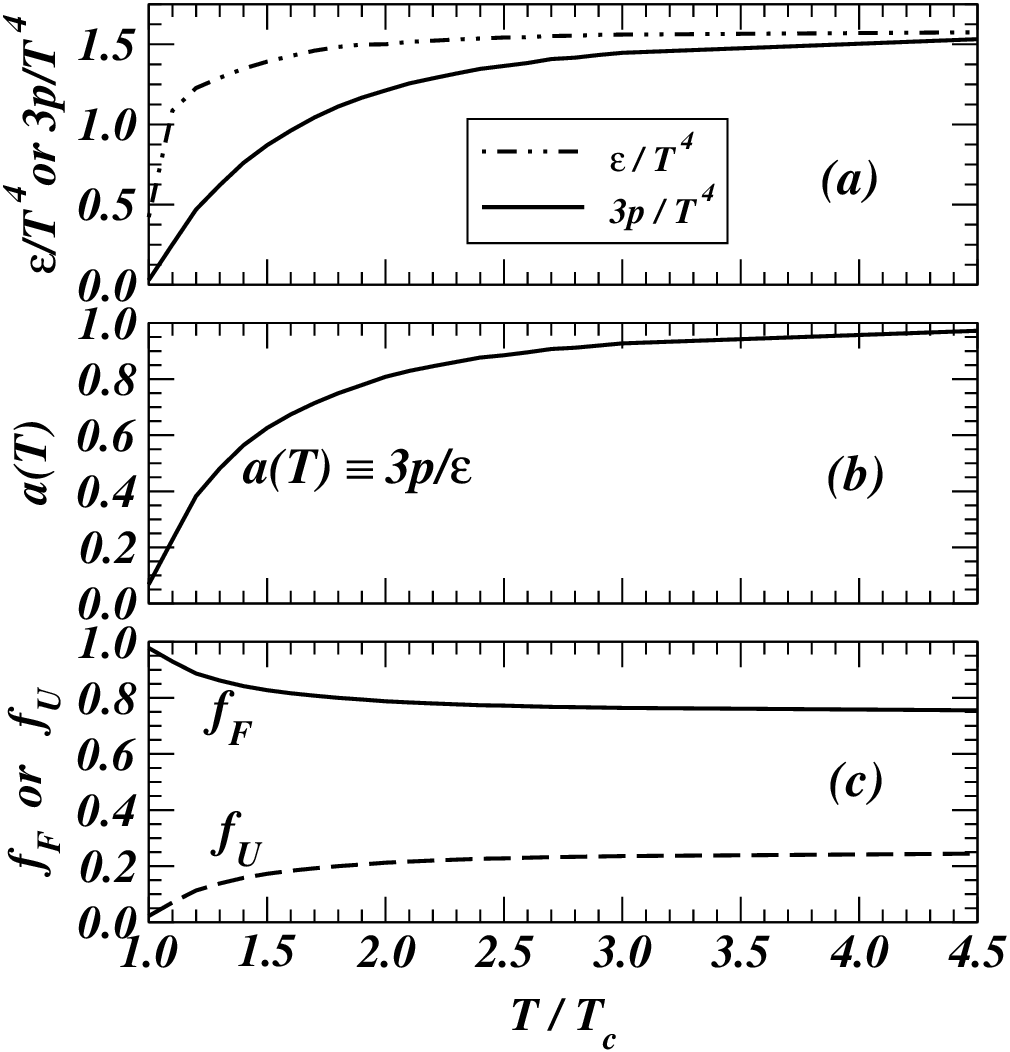}
\caption{(a) The energy density and pressure of a SU(3) gauge theory
as a function of the temperature obtained by Boyd $et~al.$
\cite{Boy96}.  (b) The ratio of $3p/\epsilon$ as a function of
$T/T_c$.  (3) The weight $f_F$ of $F_1$ and the weight $f_U$ of $U_1$
that comprise the $Q$-$\bar Q$ potential according to Eq.\
(\ref{uqqq}).  }
\end{figure}

The $Q$-$\bar Q$ potential which appears in the
Schr\" odinger equation for quarkonium bound states, Eq.\
(\ref{exact}), is then
\begin{eqnarray}
\label{uqqq}
U_{Q\bar Q}^{(1)}({\bf r},T)-
U_{Q\bar Q}^{(1)}({\bf r}\to \infty,T)=
f_F(T)
 \left \{  F_1({\bf r},T) - F_1({\bf r}\to \infty ,T) \right \}
+    f_U(T)
\left \{  U_1({\bf r},T) - U_1({\bf r}\to \infty ,T) \right \},
\end{eqnarray}
where
\begin{eqnarray}
f_F(T)=
 \frac{3}{3+a(T)},   
\end{eqnarray}
\begin{eqnarray}
f_U(T)=
 \frac{a(T)}{3+a(T)},
\end{eqnarray}
and $f_F(T)+f_F(T)=1$.  We shall use such a relation to obtain an
approximate $Q$-$\bar Q$ potential from $F_1$ and $U_1$.

The above Eq.\ (\ref{uqqq}) has been obtained for a color-singlet
$Q\bar Q$ pair in the quenched approximation.  It can be easily
generalized to the case of the unquenched full QCD with dynamical
light quarks.  In that case, the function $a(T)$ corresponds to
$p/(\epsilon/3)$ appropriate for the equation of state of the
quark-gluon plasma under consideration, and $U_{g0}(T)$ of Eq.\
(\ref{Uqq1}) becomes $U_{qgp}(T)$, the internal energy of the
quark-gluon plasma in the absence of the $Q\bar Q$ pair, and Eq.\
(\ref{uqqq}) remains unchanged.

We show $\epsilon/T^4$ and $p/T^4$ obtained in quenched QCD by Boyd
$et~al.$ \cite{Boy96} as a function of $T/T_c$ in Fig.\ 1(a).  In
Fig.\ 1(b) the function $a(T)$, defined as the ratio $3p/\epsilon$, is
plotted as a function of $T/T_c$.  The free energy fraction $f_F(T)$
and the internal energy fraction $f_U(T)$ calculated using this ratio
of $a(T)$ are shown in Fig.\ 1(c).  One finds that at temperatures
close to $T_c$, $f_F$ is close to unity, and the $Q$-$\bar Q$
potential is close to the free energy $F_1({\bf r})$.  As temperature
increases, the $F_1$ fraction decreases but approaching $f_F \sim
0.75$ at very high temperatures.  The $U_1$ fraction is nearly zero at
temperatures near $T_c$ and it increases monotonically as a function
of temperature, reaching a value of 0.25 at very high temperatures.

It is of interest to discuss the conditions under which the
application of the static potential $U_{Q\bar Q}^{(1)}({\bf r},T)$ and
its representation in terms of $F_1$ and $U_1$ as given in Eq.\
(\ref{uqqq}) can be reasonable concepts.  For the static potential and
the quark-gluon plasma equation of state to be applicable, it is
necessary that the time for the quark-gluon plasma to reach thermal
and hydrostatic equilibrium is short compared with the time for the
periodic motion of the $Q$ and $\bar Q$.  The time for the quark-gluon
plasma to reach thermal equilibrium is of the order 0.6 fm/c, as one
may infer from discussions in Section II.  The orbiting time for a
heavy quarkonium is of order $2r_{\rm rms}/v$ where $r_{\rm rms}$ is
the root-mean-square radius of the heavy quarkonium system and $v$ is
the relative velocity which is at most of order 0.5 for heavy quarks.
The spatial scale of the heavy quarkonia in the quark gluon plasma is
quite large.  As we shall see in Section VIII and IX $r_{\rm rms}=
0.88$ fm at $T/T_c=1.13$ for $J/\psi$ and it increases to $r_{\rm
rms}=5.3$ fm at $T/T_c=1.65$.  For $\Upsilon$, $r_{\rm rms}= 0.25$ fm
at $T/T_c=1.13$ and it increases to $r_{\rm rms}=0.59$ fm at
$T/T_c=2.6$.  The large spatial scales arises because these heavy
quarkonium states are basically only weakly bound.  The orbiting time
for heavy quarkonia in the quark-gluon plasma is much greater than the
quark-gluon plasma thermalization time.  It is therefore reasonable to
use a static $Q$-$\bar Q$ potential and the equation of state of the
quark-gluon plasma to study heavy quarkonium in the quark-gluon
plasma.

\section{Simple Parameterizations of $U_1$ and $F_1$}

Kaczmarek $et~al.$ \cite{Kac03} obtained the color-singlet free energy
$F_1({\bf r},T)$ and internal potential $U_1({\bf r},T)$ in quenched
QCD as a function of $r=|{\bf r}|$ and $T$.  The radial dependences of
$F_1({\bf r},T)$ and $U_1({\bf r},T)$ in quenched QCD can be
adequately represented by a screened Coulomb potential with a
screening mass $\mu_i$, a coupling constant $\alpha_i$, and an
asymptotic potential horizon $C_i(T)$ at $|{\bf r}| \to \infty$, where
the subscript $i=F$ and $U$ stands for the free energy or the internal
energy respectively,
\begin{eqnarray}
\label{scoulF}
F_1({\bf r},T)=C_F(T)-\frac{4\alpha_F(T)}{3} \frac{e^{-\mu_F r}}{r},
\end{eqnarray}
and 
\begin{eqnarray}
\label{scoulU}
U_1({\bf r},T)=C_U(T)-\frac{4\alpha_U(T)}{3} \frac{e^{-\mu_U r}}{r}.
\end{eqnarray}

\begin{figure} [h]
\includegraphics[angle=0,scale=0.50]{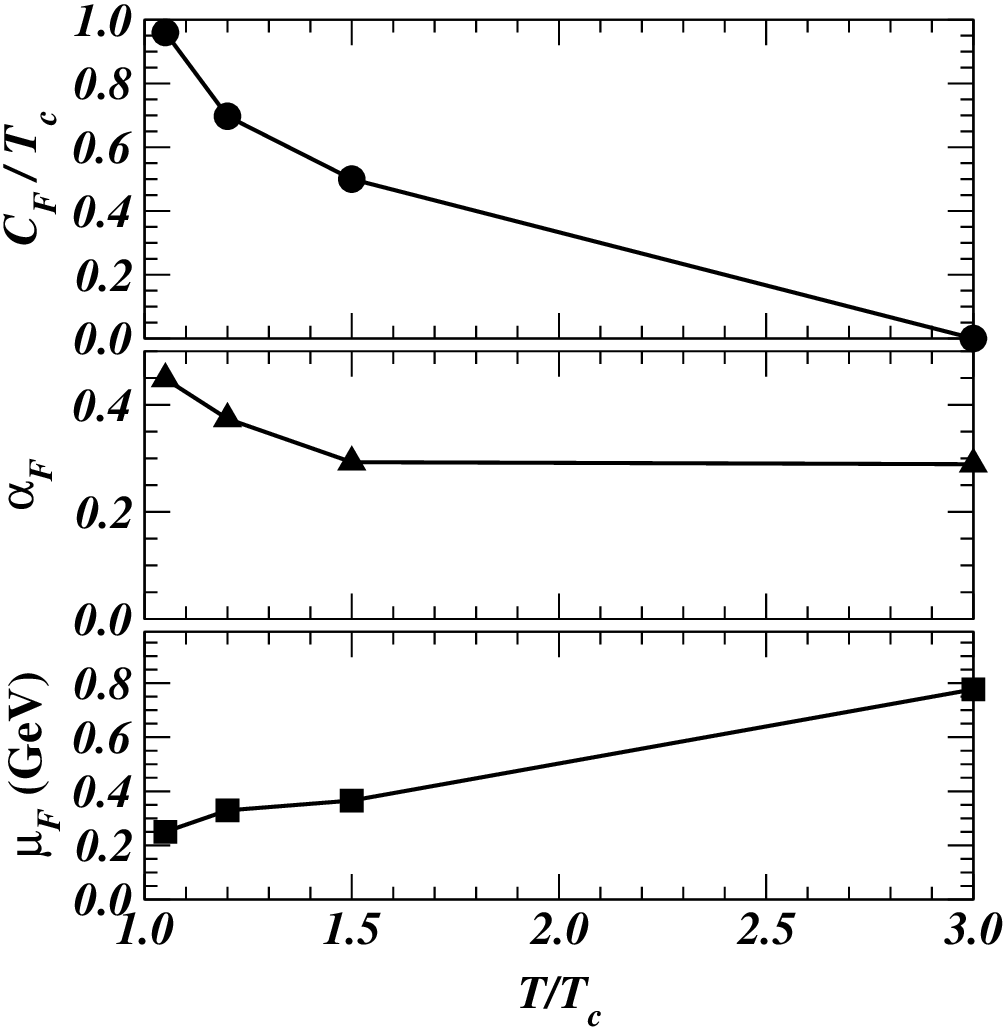}
\caption{The color-singlet parameters $C_F$, $\alpha_F$, and $\mu_F$ for the
free energy $F_1({\bf r},T)$ as given in Eq.\ (\ref{scoulF}).}
\end{figure}

The parameters $C_i$, $\alpha_{i}$, and $\mu_i$ for the quenched
lattice QCD results of $F_1({\bf r},T)$ and $U_1({\bf r},T)$ of
Kaczmarek $et~al.$ \cite{Kac03} are shown in Figs.\ 2 and 3, and the
corresponding fits to the lattice $F_1$ and $U_1$ results are shown in
Figs. 4 and 5.  The coupling constant $\alpha_{F}(T)$ is 0.44 at
$T=1.05\, T_c$.  As the temperature increases, $\alpha_F$ decreases
and saturates at $\alpha_F \sim 0.3$ at $T \sim 3 T_c$.  The screening
mass $\mu_F$ is about 0.25 GeV at temperatures just above $T_c$ and it
increases to 0.8 GeV at $3T_c$.
\begin{figure} [h]
\includegraphics[angle=0,scale=0.50]{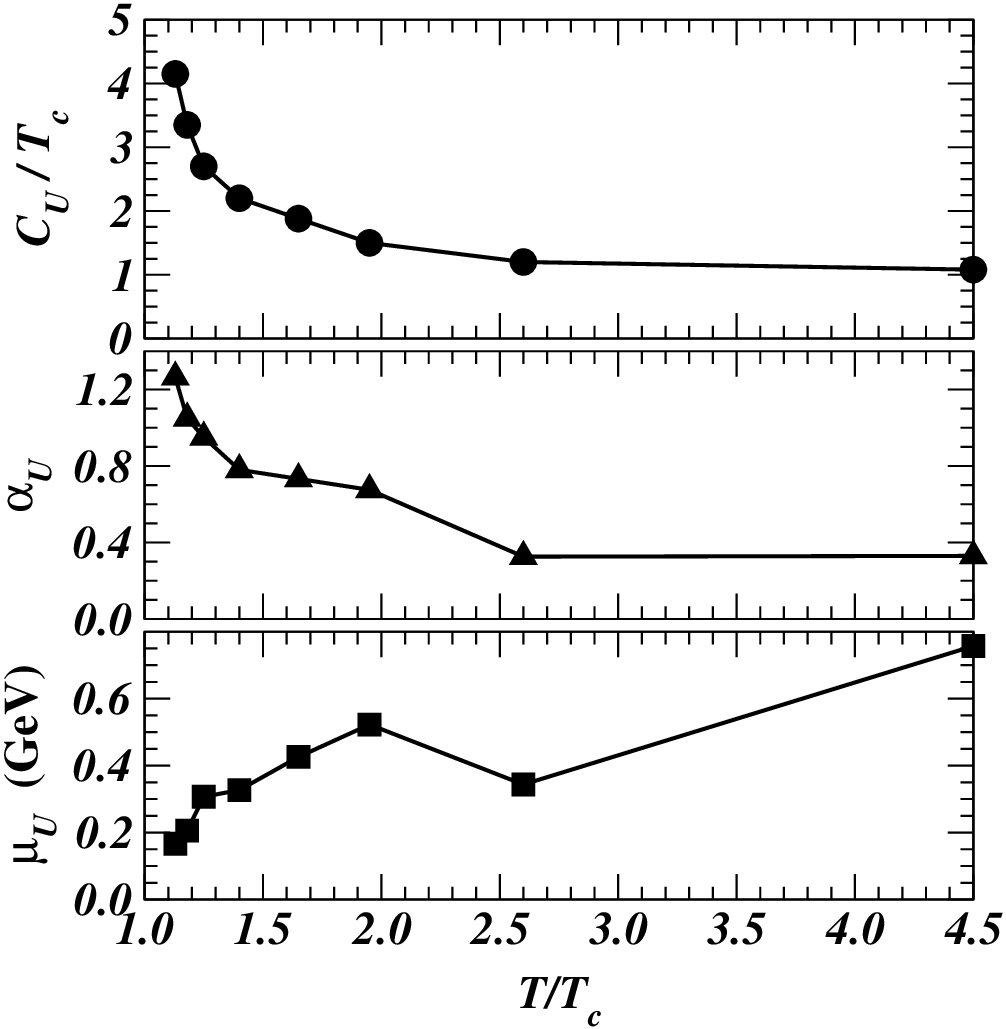}
\caption{The color-singlet parameters $C_U$, $\alpha_U$, and $\mu_U$ 
for the total internal energy $U_1({\bf r},T)$ as given in Eq.\
(\ref{scoulU}).}
\end{figure}
The coupling constant $\alpha_{U}(T)$ is quite large at temperatures
slightly above the phase transition temperature.  At $T=1.13\, T_c,
\alpha_U=1.26$.  As the temperature increases, $\alpha_U$ decreases
and saturates at $\alpha_{U}(T) \sim 0.4$ at $T \sim 4 T_c$.  The
screening mass $\mu_U$ is small at temperatures just above $T_c$.  As
the temperature increases, the screening mass $\mu_U$ increases to
about 0.8 GeV at $T\sim 4.5 T_c$.

\vspace*{-0.0cm}
\hspace*{1.5cm}
\begin{figure} [h]
\includegraphics[angle=0,scale=0.45]{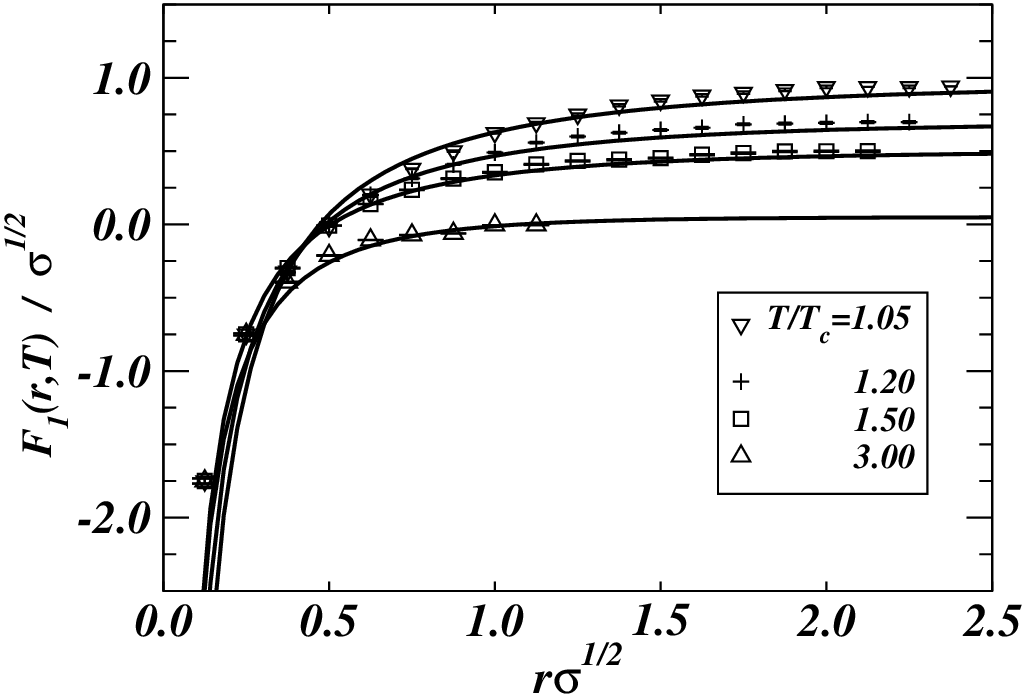}
\caption{The symbols represent the quenched lattice QCD free energy,
$F_1(r,T)/\sigma^{1/2}$, of Kaczmarek $et~al.$ \cite{Kac03} at
selected distances and the curves are the fits using the screened
Coulomb potential, Eq.\ (\ref{scoulF}), with parameters given in
Fig. 2. Here $\sigma^{1/2}=425 $ MeV.}
\end{figure}

The comparison in Fig. 4 and 5 shows that the free energy $F_1$ and
the internal energy $U_1$ with the set of parameters in Figs.\ 2 and
3, adequately describe the lattice-gauge data and can be used to
calculate the eigenvalues and eigenfunctions of heavy quarkonia.  In
the local energy-density approximation, the $Q$-$\bar Q$ potential is
given by
\begin{eqnarray}
\label{finalU}
  U_{Q\bar Q}^{(1)}({\bf r},T) 
- U_{Q\bar Q}^{(1)}({\bf r}\to \infty,T) 
= -\frac{4}{3} \frac{      f_F(T)~ \alpha_F(T) e^{-\mu_F r} 
                     +  f_U(T)~ \alpha_U(T) e^{-\mu_U r} } { r }.
\end{eqnarray}

\vspace*{-0.0cm}
\hspace*{1.5cm}
\begin{figure} [h]
\includegraphics[angle=0,scale=0.45]{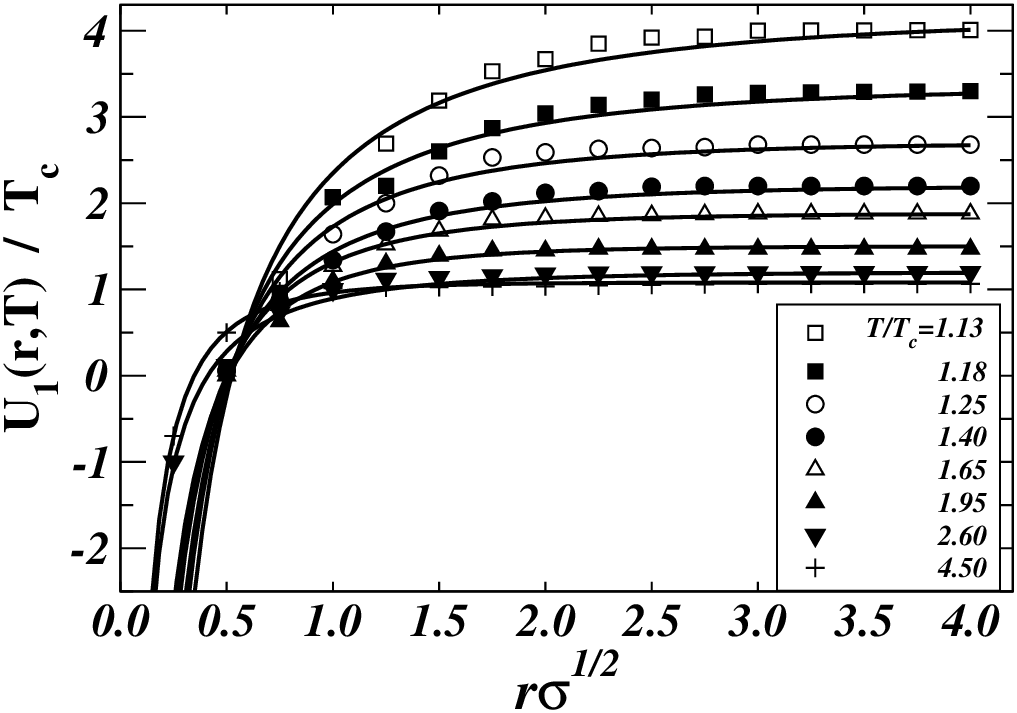}
\caption{The symbols represent the quenched lattice QCD total internal energy, 
$U_1(r,T)/T_c$, of Kaczmarek $et~al.$ \cite{Kac03} at selected
distances and the curves are the fits using the screened Coulomb
potential, Eq.\ (\ref{scoulU}), with parameters given in Fig. 3. Here
$T_c=269 $ MeV and $\sigma^{1/2}=425 $ MeV.}
\end{figure}

\section{Charmonium in the Quark-Gluon Plasma}

In the quenched approximation, the transition temperature is $T_c=269$
MeV \cite{Kar03}.  We use this value of $T_c$ to express the potential
in GeV units, in order to evaluate the energy levels of different
heavy quarkonia.

For a given temperature, we use the $U_{Q\bar Q}^{(1)}({\bf r},T)$
potential given in Eq.\ (\ref{uqqq}) to calculate the charmonium
energy levels and wave functions.  In these calculations, we employ a
charm quark mass $m_c= 1.3$ and 1.5 GeV
\cite{Hag02} to provide an indication of the uncertainties of 
the eigenenergies.

\begin{figure}[h]
\hspace*{0.7cm}
\includegraphics[angle=0,scale=0.50]{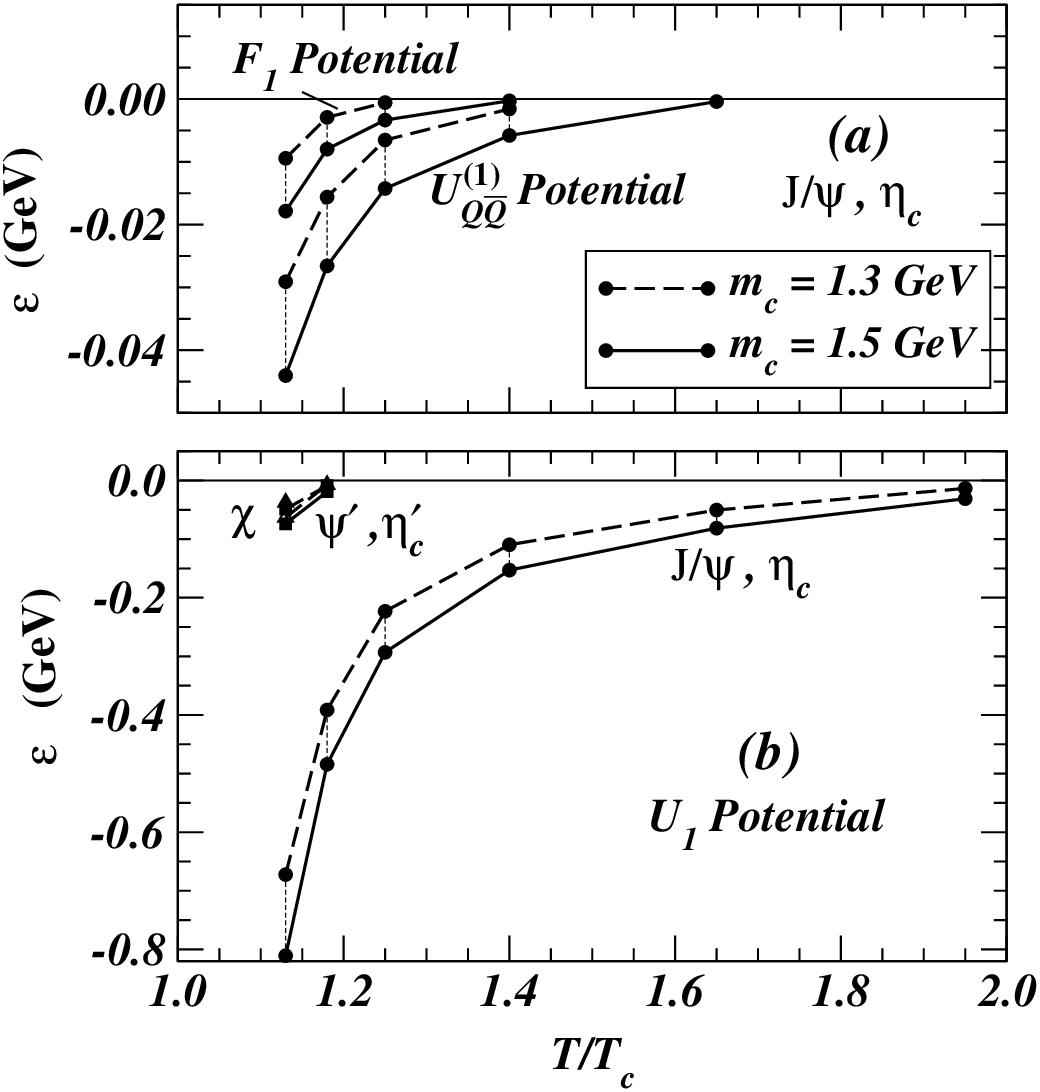} 
\caption{Energy levels of charmonium in the quark-gluon plasma as a
function of temperature calculated with ($a$)  the $F_1({\bf r},T)$ and
$U_{Q\bar Q}^{(1)}({\bf r},T)$ potentials for $J/\psi$ and $\eta_c$,
and ($b$) the $U_1({\bf r},T)$ potential for $J/\psi, \eta_c,
\chi_c, \psi',$ and $\eta_c'$.  The dashed curves are obtained with
$m_c=1.3$ GeV and the solid curves with $m_c=1.5$ GeV. Note the
difference in the enegy scales in ($a$) and ($b$).}
\end{figure}

Energy levels of charmonium calculated with the $U_{Q\bar
Q}^{(1)}({\bf r},T)$ potential are shown in Fig.\ 6 as a function of
temperature.  The $J/\psi$ and $\eta_c$ states are weakly bound at
temperatures slightly greater than $T_c$.  The eigenenergies of
$J/\psi$ and $\eta_c$ are -0.045 GeV at $T=1.13\, T_c$ for $m_c=1.5 $
GeV and their energies increase as the temperature increases.  The
$J/\psi$ and $\eta_c$ state eigenenergies are -0.0004 GeV at $T=1.65\,
T_c$ for $m_c=1.5$ GeV.  If one extrapolates the eigenenergy from
lower temperature points, one infers that the $J/\psi$ and $\eta_c$
spontaneous dissociation temperature is $1.52\,T_c$ for $m_c=1.3$ GeV,
and is $1.72\,T_c$ for $m_c=1.5$ GeV, with a mean value of
$1.62\,T_c$.  There are no bound $\chi_c$, $\psi'$, and $\eta_c'$
states for temperatures above $T_c$.

\begin{figure}[h]
\includegraphics[angle=0,scale=0.55]{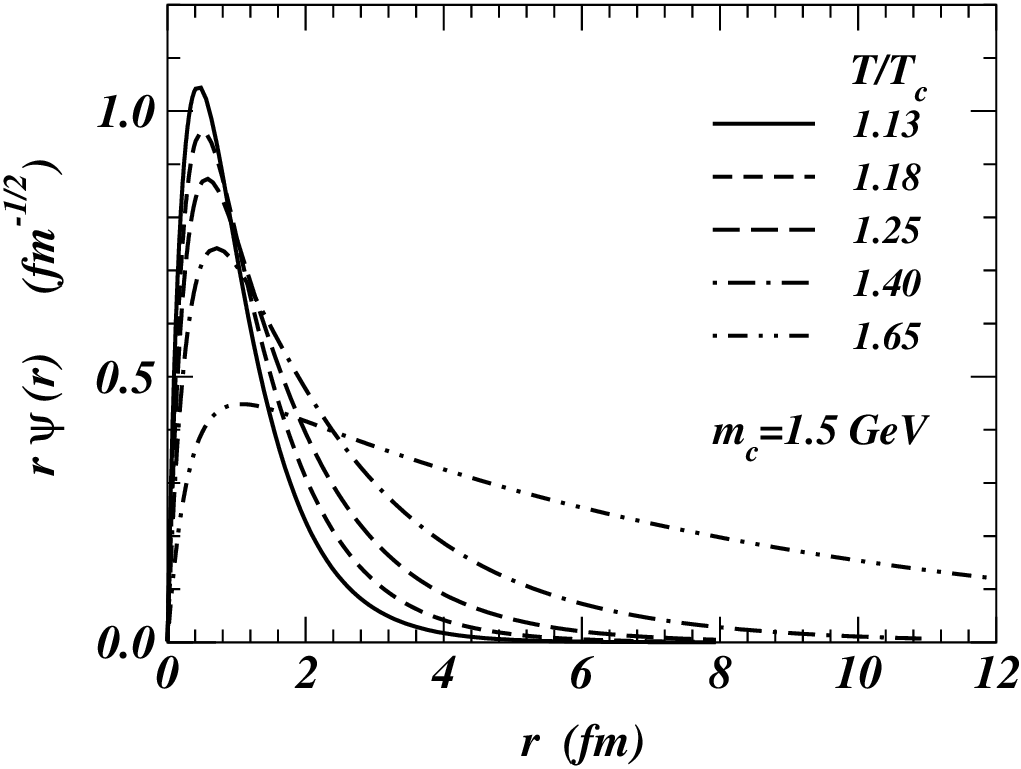}
\caption{$J/\psi$ wave function calculated with the $U_{Q\bar Q}^{(1)}({\bf r},T)$
potential for different temperatures.}
\end{figure}

In the spectral function analyses of Asakawa $et~al.$
\cite{Asa03,Asa04} and Petreczky $et~al.$ \cite{Pet03,Dat03,Pet04},
the widths of $J/\psi$ begin to be broadened at $\sim 1.6\, T_c$ and
the $\chi_c$ states are found to be dissolved already at $1.13\, T_c$.
The width of $J/\psi$ can be broadened by gluon dissociation $g+J/\psi
\to c + \bar c$ which is presumably a possible process in the lattice
gauge calculations in the spectral function analysis.  The gluon
dissociation width is however of the order of 0.05 to 0.1 GeV as one
can infer later from Section XII.  In the numerical results of the
spectral function analysis \cite{Asa03,Asa04}, the width appears to be
broadened by an amount significantly greater than this amount for
gluon dissociation.  It is therefore reasonable to associate the
broadening of the width of a heavy quarkonium in the spectral function
analysis with the occurrence of spontaneous dissociation, when the
heavy quarkonium becomes unbound.  The temperature at which the width
of a quarkonium begins to broaden significantly can be taken as the
dissociation temperature for spontaneous dissociation of the
quarkonium in the spectral function analysis.

In Table I we list the dissociation temperatures of different
quarkonia obtained in quenched QCD.  A comparison of the dissociation
temperatures from the spectral function analysis
\cite{Asa03,Asa04,Pet03,Dat03,Pet04} and from the
$U_{Q\bar Q}^{(1)}({\bf r},T)$ potential model indicates the agreement
that $J/\psi$ is bound up to about $1.6 T_c$ and is unstable at very
high temperatures.  There is also the agreement that the $\chi_c$ and
$\psi'$ states are unbound in the quark-gluon plasma.

It is instructive to compare the eigenenergies obtained by using
other potential models.  We calculate the heavy quarkonium
eigenenergies with the $F_1({\bf r},T)$ potential as in
\cite{Dig01a,Dig01b,Won01a,Won02} by replacing $U_{Q\bar Q}^{(1)}$ in
Eq.\ (\ref{exact}) with $F_1({\bf r},T)$.  In Fig.\ 7(a), we show the
charmonium energies calculated with the $F_1({\bf r},T)$ potential.
One finds that $J/\psi$ is weakly bound, but the dissociation
temperature lies in the range 1.33-1.46 $T_c$ for $m_c=1.3-1.5$ GeV,
with a mean value of $1.40\, T_c$ .  This dissociation temperature is
lower than that inferred from the spectral function analysis.  The
$\chi_c$ and $\psi'$ states are unbound in this potential.

We also calculate the heavy quarkonium eigenenergies with the total
internal energy $U_1({\bf r},T)$ as the $Q$-$\bar Q$ potential as in
\cite{Dig01a,Dig01b,Won01a,Won02}. The eigenenergies for charmonium states
are shown in Fig.\ 7(b). As the total internal energy $U_1$ contains a
deeper potential well, the charmonium states are deeply bound.  The
binding energy is about 0.8 GeV at $1.13\, T_c$, and the state becomes
unbound at 2.50-2.71 $T_c$, with a mean value of $2.60\, T_c$.  This
dissociation temperature is much higher than dissociation temperature
of about 1.6 $T_c$ obtained from the spectral function analysis.
There are uncertainties in the spontaneous dissociation temperatures
due to the differences in the degrees of freedom assumed in lattice
QCD calculations.  For example, using the total internal energy $U_1$
extracted from the full QCD with two flavors obtained by Kaczmarek
$et~al.$ \cite{Kac03b}, Shuryak found that the dissociation
temperature of $J/\psi$ is about 2.7 $T_c$
\cite{Shu04a}.
In this model, the $\chi_c$, $\psi'$ and $\eta_c'$ states are bound at
temperature slightly above $T_c$ and they become unbound at 1.2 $T_c$.
The binding of $\chi_c$ states is in disagreement with the results of
the dissolution of $\chi_c$ states in the spectral function analysis.

Our comparison of these results indicate that the model that
compares best with the spectral function analysis is the $U_{Q\bar
Q}^{(1)}({\bf r},T)$ potential [Eq.\ ({\ref{uqqq})] obtained from a
variational principle and the quark-gluon plasma equation of state.
However, as the results from the spectral function analysis are still
scanty, more results from the spectral function analysis are needed to
test further the potential model.

\vskip 0.5cm \centerline{Table I.  Dissociation temperatures 
obtained from different analyses in quenched QCD.} 
{\vskip 0.5cm\hskip 1.5cm
\begin{tabular}{|c|c|c|c|c|}
\hline
{\rm Heavy Quarkonium} & ~ $U_{Q\bar Q}^{(1)}({\bf r},T)$ {\rm Potential}
                       & ~ $F_1({\bf r},T)$ {\rm Potential} 
                       & ~ $U_1({\bf r},T)$ {\rm Potential}
                       & ~{\rm Spectral Analysis}~ \\ 
\hline
$J/\psi,\eta_c$       &   $1.62\,T_c$ &   $1.40\,T_c$  
&   $2.60 \,T_c$ & $\sim 1.6 \,T_c$              \\ \cline{1-5}  
$\chi_c$              &   unbound in QGP         &   unbound in QGP 
&   $1.19 \,T_c$      & dissolved below $ 1.1\, T_c$  \\ \cline{1-5}
$\psi',\eta_c'$       &   unbound in QGP          &   unbound in QGP 
&   $ 1.20 \,T_c$      &                               \\ \cline{1-5}
$\Upsilon,\eta_b$     &    $ 4.10 \,T_c$    &    $ 3.50 \,T_c$   
&   $\sim 5.0 \,T_c$  &                               \\ \cline{1-5}  
$\chi_b$              & $ 1.18 \,T_c$      &   $ 1.10 \,T_c$ 
& $ 1.73 \,T_c$       &                               \\ \cline{1-5}
$\Upsilon',\eta_b'$   & $ 1.38 \,T_c$      & $ 1.19 \,T_c$         
& $2.28 \,T_c$        &                               \\ \cline{1-5}
                                                        \hline
\end{tabular}
}
\vskip 0.6cm

The solution of the Schr\"odinger equation (\ref{sch}) gives both
eigenenergies and eigenfunctions.  We show in Fig. 8 the wave
functions of $J/\psi$ calculated with the $U_{Q\bar Q}^{(1)}$
potential and $m_c=1.5$ GeV.  They are normalized according to
\begin{eqnarray}
\label{nor}
\int_0^\infty |u_{1S}(r)|^2 dr =
\int_0^\infty |r \psi_{1S}(r)|^2 dr =
1, 
\end{eqnarray}
as in Eq.\ (4.18) of Blatt and Weisskopf \cite{Bla52}.  As the
temperature increases, the binding of the state becomes weaker and the
wave function extends to greater distances.  The rms $r$ of the
$J/\psi$ wave function is 0.88 fm at $1.13 T_c$.  At $T=1.65 \,T_c$,
which is near the temperature for spontaneous dissociation, the
binding energy is 0.0004 GeV.  The rms $r$ of the $J/\psi$ wave
function is 5.30 fm, which is much greater than the theoretical rms
$r$ of 0.404 fm for $J/\psi$ at zero temperature \cite{Won01}.

\section{   Bottomium bound states in the Quark-Gluon Plasma}

\hspace*{0.0cm}
\begin{figure}[h]
\includegraphics[angle=0,scale=0.50]{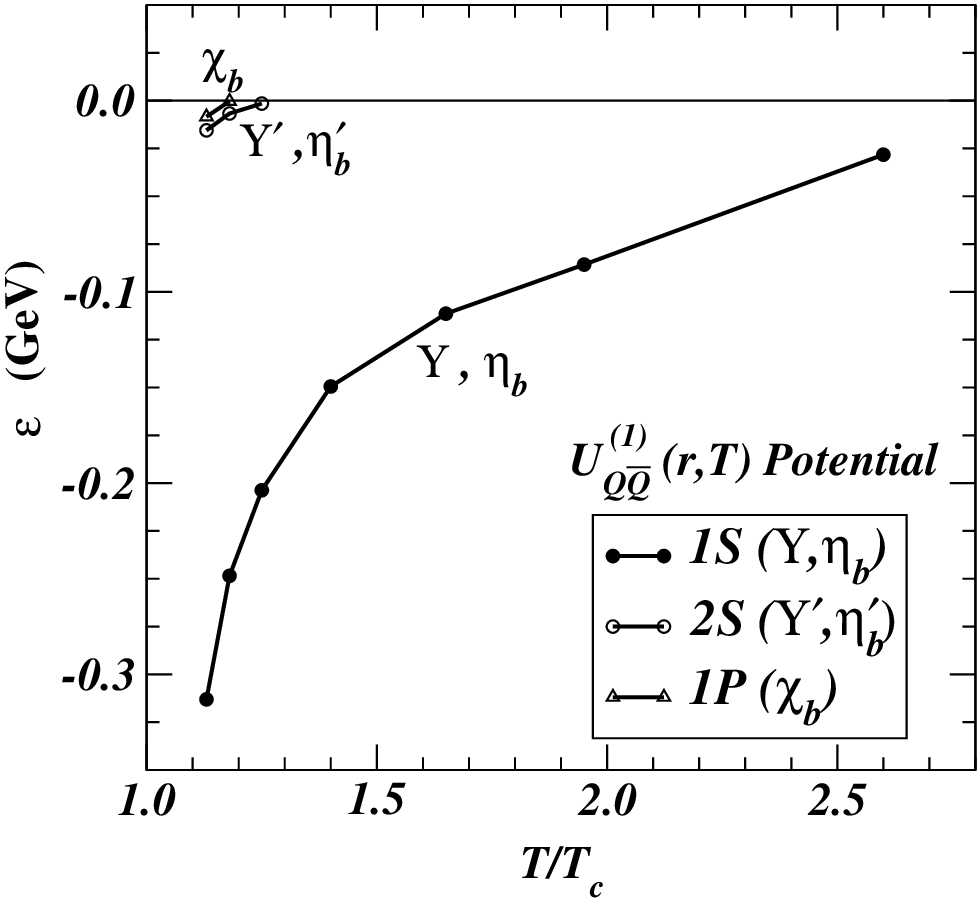}
\caption{Energy levels of $b\bar b$ bound states as a function of
temperature, calculated with the $U_{Q\bar Q}^{(1)}({\bf r},T)$
potential.}
\end{figure}

One can carry out similar calculations for the energy levels and wave
functions of the $b$-$\bar b$ system.  We take the mass of the bottom
quark to be 4.3 GeV, which falls within the range of 4.1 to 4.5 GeV in
the PDG listing \cite{Hag02}.  The energy levels of the lowest $b\bar
b$ bound states calculated with the $U_{Q\bar Q}^{(1)}({\bf r},T)$
potential, are shown in Fig. 9 as a function of temperature.  We find
that at $T = 1.13 \,T_c$, the $\Upsilon$ state lies at about -0.3 GeV
and the state energy increases as the temperature increases.  The
$\Upsilon$ state remains to be bound by 0.028 GeV at $T=2.5 \, T_c$.
If one extrapolates from lower temperatures, the dissociation
temperature of $\Upsilon$ and $\eta_b$ is $4.10\, T_c$.  For this
potential, the $\chi_b$, $\Upsilon'$, and $\eta_b'$ states are weakly
bound at temperatures slightly greater than $T_c$.  The $\chi_b$ states
become unbound at $1.18\, T_c$ and the $\Upsilon'$ and $\eta_b'$ become
unbound at $1.38 T_c$.

\hspace*{0.0cm}
\begin{figure}[h]
\includegraphics[angle=0,scale=0.50]{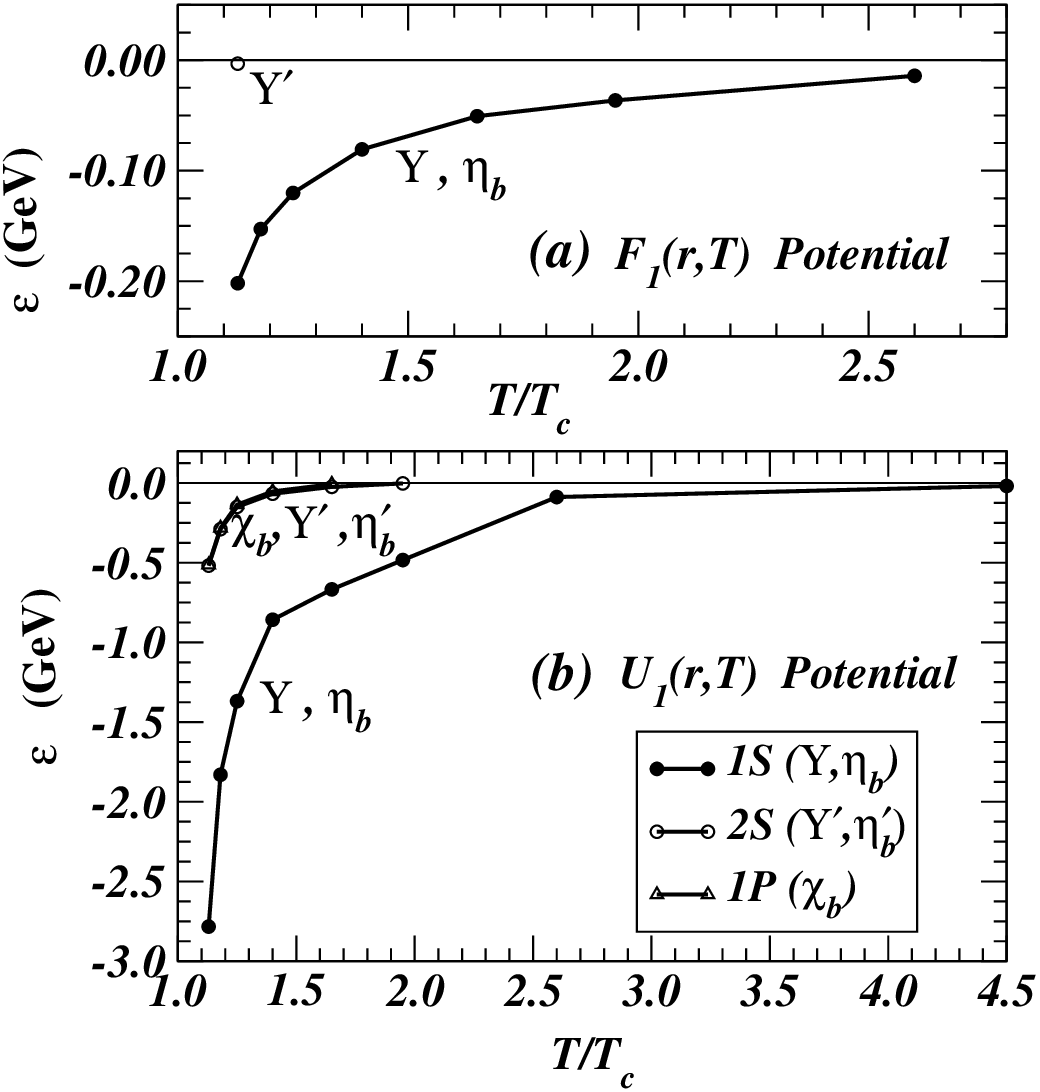}
\caption{Energy levels of $b\bar b$ bound states as a function of
temperature calculated with (a)the $F_1({\bf r},T)$ potential and (b)
the $U_1({\bf r},T)$ potential.}
\end{figure}

As a comparison, we calculate the bottomium eigenenergies using other
potential models.  If we assume that the $Q$-$\bar Q$ potential is the
free energy, $F_1({\bf r},T)$, we find that $\Upsilon$ is bound by
about 0.2 GeV at $1.13\, T_c$ and it becomes unbound at $3.50 T_c$ as
shown in Fig.\ 10(a).  The $\chi_b$ states are unbound at $1.10 T_c$,
and the $\Upsilon'$ and $\eta_b'$ states are bound by .003 GeV at
$1.13 T_c$, which should be close to its dissociation temperature.

If we assume that the $Q$-$\bar Q$ potential is the total internal
energy, $U_1({\bf r},T)$, we find that the bottomium states become
deeply bound as shown in Fig.\ 10(b).  At 1.13 $T_c$, the $\Upsilon$
and $\eta_b$ states are bound by about 3 GeV.  The binding energy
decreases slowly as a function of temperature.  The small binding
energy at $T=4.5\,T_c$ indicates that the $\Upsilon$ dissociation
temperature is close to and slightly greater than $T\sim 5.0 \, T_c$.
Because of the small screening mass near $T_c$, the potential for
temperatures near $T_c$ is approximately a Coulomb potential but with
a large coupling constant.  Hence, $\chi_b$ and $\Upsilon', \eta_b'$
states are nearly degenerate.  They lie at -0.5 GeV for $T=1.13 \,
T_c$, and begin to be unbound at $1.73\,T_c$ and $2.28\, T_c$
respectively.

The $b\bar b$ bound state wave functions can be obtained by solving
the Schr\"odinger equation (\ref{sch}).  We show in Fig.\ 11 the
$\Upsilon$ radial wave functions as a function of temperature.  The
wave function extends to greater distances as the temperature
increases.  The root-mean-square radius $r_{\rm rms}$ is 0.25 fm at
$T/T_c=1.13$ and it increases as the temperature increases.  At $T=
2.6 \, T_c$, which lies very close to the temperature for spontaneous
dissociation, the rms $r$ value is 0.59 fm, which is substantially
greater than the rms $r$ of 0.25 fm at $1.13\, T_c$.

\hspace*{0.0cm}
\begin{figure} [h]
\label{wf}
\includegraphics[angle=0,scale=0.50]{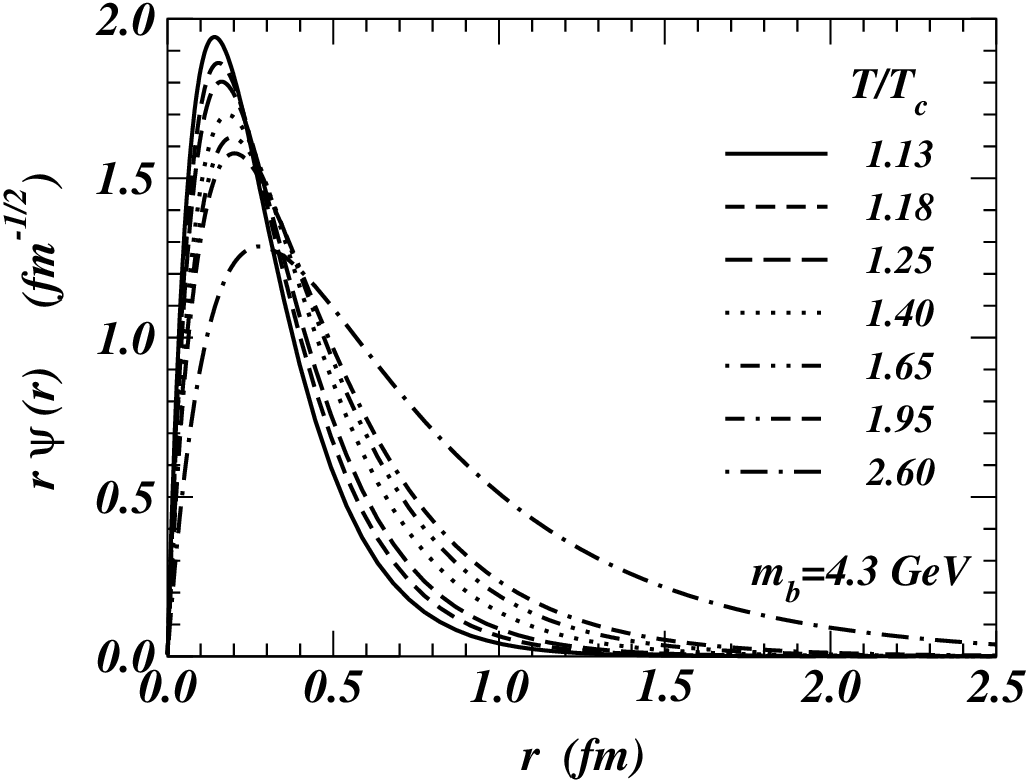}
\caption{$\Upsilon$ wave function $r \psi(r)$ as a function of
temperature.}
\end{figure}

Our comparisons of the charmonium eigenenergies with those from the
spectral function analysis indicate that the results obtained by using
the $U_{Q\bar Q}^{(1)}$ potential give results which agree best with
the spectral function analysis.  As different models give different
predications on the bottomium dissociation temperatures, further tests
of the models can be carried out by calculating the dissociation
temperature of bottomium states using the spectral function analysis.

\section{Antiscreening by deconfined gluons} 

As one crosses the phase transition temperature $T_c$ from below,
quarks and gluons becomes deconfined. The Debye screening due to the
interaction between the heavy quark and antiquark with medium
particles is considered to be the dominant effect when a heavy
quarkonium is placed in a quark-gluon plasma \cite{Mat86}.  It is
argued that Debye screening leads to a decrease in the attractive
interaction between the heavy quark and antiquark and results in the
spontaneous dissociation of the heavy quarkonium in the quark-gluon
plasma.  The suppression of $J/\psi$ production was suggested as a
signature for the quark-gluon plasma \cite{Mat86}.  

In perturbative QCD, there is a relation between the screening mass
$\mu$, coupling constant $g=\sqrt{4 \pi \alpha_s}$, and temperature
$T$ in a quark-gluon plasma given by
\cite{Kli81,Wel82,Kap89,Bra92,Reb92}
\begin{eqnarray}
\mu^2 = g^2 T^2 (N_c+N_f/2)/3.
\end{eqnarray}
For the quenched approximation with $N_f=0$, perturbative QCD gives
$\mu=gT$.  A comparison of the effective coupling constants $\alpha_F$
and $\alpha_U$ of either $F_1$ or $U_1$ with the screening mass
$\mu_F$ and $\mu_U$ at temperatures near $T_c$ indicates that the
screening mass is much smaller than the estimates in perturbative QCD
limit, indicating that the large-distance behavior cannot be
qualitatively described by perturbative QCD. The perturbative QCD
limit for the large-distance behavior can be reached only at
temperatures above 6 $T_c$ \cite{Kar01a,Kac04}.  In fact, Kaczmarek
$et~al.$ \cite{Kac04} pointed out that for temperatures close to $T_c$
the QCD phase should be more appropriately described in terms of
remnants of the confinement part of QCD rather than a strongly coupled
Coloumbic force \cite{Shu03,Shu04a}.  To understand this ``remnant of
QCD confinement above $T_c$'', it is of interest to examine effects of
antiscreening above $T_c$ by deconfined gluons.

In a related topic, Svetitsky, Yaffe, DeTar and DeGrand previously
found that large space-like Wilson loops in the quark-gluon plasma
have an area law behavior \cite{Sve82,Yaf82,Det85,Deg86,Det88}.  As
the area law of the spatial Polyakov loop does not change drastically
across the phase transition, DeTar \cite{Det85,Det88}, Hansson, Lee,
and Zahed \cite{Han88}, and Simonov \cite{Sim95,Sim95a,Sim05} argued
that low-lying mesons including $J/\psi$ may remain relatively narrow
states due to the attractive interaction between the quark and the
antiquark and the suppression of $J/\psi$ is not a signature of the
deconfinement phase transition \cite{Det88}.  It is of interest to
examine the consequence of the spatial area law to see whether it will
lead to antiscreening between a heavy quark and antiquark.

The mechanism of antiscreening by virtual gluons at $T=0$ is well
known (see for example  Peskin, Schroeder, Gottfried, and Weisskopf
\cite{Got86,Pes96}).  We would like to follow similar arguments
to study the mechanism of antiscreening by deconfined gluons above
the phase transition temperature $T_c$.  We consider the color
electric field generated by a static color source $\rho_{\rm
ext}^{a(0)}({\bf r})= \delta({\bf r})\delta^{a
\lambda} $ with a unit color charge of index $\lambda$ placed at the
origin, in the presence of an external gauge field (Fig.\ 12$a$).  We
fix the gauge to be the Coulomb gauge and represent the deconfined
gluons in terms of an external transverse gluon field $A^{bi}({\bf
r})$ where the first index $b=1,...,8$ is the color index and the
second index $i=1,2,3$ is the spatial coordinate index. 

The color electric field $E^{ai}({\bf r})$ generated by the source is
determined by the Gauss law
\begin{eqnarray}
\label{nonl}
\partial_i E^{ai}({\bf r}) = g \delta({\bf r}) \delta^{a \lambda}  + g 
f^{abc} A^{bi}({\bf r}) E^{ci}({\bf r})
\end{eqnarray}
(Eq. (16.139) of Peskin and Schroeder \cite{Pes96}).  Here repeated
indices are summed over and the first index of $E^{ai}({\bf r})$ is
the color index and the second index is the spatial coordinate index.
Because of the non-linear nature of the second term which arises from
the non-Abelian nature of QCD, the external color source $\delta({\bf
r}) \delta^{a \lambda}$ and the external gauge field $A^{ai}({\bf r})$
induce a color source $\rho_{\rm ind}^{a(1)}({\bf r})$, which in turn
induces an additional color source $\rho_{\rm ind}^{a(2)}({\bf r})$.
How do these induced color charges depend on the external gauge field
$A^{bi}({\bf r})$?  

We consider an expansion of the source in terms of the external source
and the induced sources, in powers of the coupling constant
\begin{eqnarray}
\partial_i E^{ai}({\bf r}) = g \delta({\bf x}) \delta^{a1}  
+ g \rho_{\rm ind}^{a(1)}({\bf r}) +  g \rho_{\rm ind}^{a(2)}({\bf r}) + ...
\end{eqnarray}
In the Coulomb gauge, the color field $E^{ci(1)}({\bf r})$, as arising
only from the external static source $\delta({\bf r})\delta^{a\lambda} $, is
\begin{eqnarray}
E^{a i(1)}({\bf r})=g \delta^{a\lambda}\frac{ r^i}{r^3}. 
\end{eqnarray}
From the non-linear term in Eq.\ (\ref{nonl}), the color charge
density induced at ${\bf r}$ by the external gauge field $A^{\beta
i}({\bf r})$ and the electric field $E^{ci(1)}({\bf r})$ of the
external color source is
\begin{eqnarray}
\label{first}
\rho_{\rm ind}^{a(1)} ({\bf r}) =  f^{a \beta \gamma} A^{\beta i}({\bf r})
E^{\gamma i(1)}({\bf r})
= g f^{a \beta \gamma} A^{\beta i}({\bf r})
\frac{\delta^{\gamma \lambda}  r^i }{r^3}.
\end{eqnarray}
An induced color-charge element $\rho_{\rm ind}^{a(1)} ({\bf r})
\Delta {\bf r}$ at ${\bf r}$ will generate a field $E^{a i (2)}({\bf
r}',{\bf r})$ at ${\bf r }'$ and this field is pointing along the
direction of ${\bf r}'-{\bf r}$,
\begin{eqnarray}
E^{a i (2)}({\bf r}',{\bf r})=g  [\rho_{\rm ind}^{a(1)}({\bf r}) \Delta {\bf r}] 
\frac{({ r}'^{i}-{ r}^i)}{|{\bf r}' -{\bf r}|^3}.
\end{eqnarray}
From the non-linear term in Eq.\ (\ref{nonl}), the color charge
density element $\rho_{\rm ind}^{a(2)} ({\bf r'},{\bf r}) \Delta{\bf
r}$, that is induced at ${\bf r}'$ by the external gauge field $A^{b
i}({\bf r}')$ and the electric field $E^{c i (2)}({\bf r'} ,{\bf r})$
at ${\bf r}'$, is therefore
\begin{eqnarray}
\label{second}
\rho_{\rm ind}^{a(2)}({\bf r}',{\bf r}) \Delta {\bf r}=  f^{a b c}  A^{bi} ({\bf r}') 
E^{c i (2)}({\bf r}',{\bf r})
=
g^2 f^{a b c}  A^{bi}({\bf r}')  
\left [ f^{c \beta \gamma} A^{\beta j} ({\bf r}) 
\frac{ \delta^{\gamma \lambda} { r}^j }{|{\bf r}|^3} \Delta{\bf r} \right ]
\frac{({ r}^{'i}-{ r}^i)}{|{\bf r}' -{\bf r}|^3} .
\end{eqnarray}
As the external source has the color index $\lambda$, we would like to
study the induced color charge of index $\lambda$ to see whether the
induced color charges lead to screening or antiscreening.  From Eq.\
(\ref{first}), we have
\begin{eqnarray}
\rho_{\rm ind}^{\lambda (1)} ({\bf r}) = 0 ,
\end{eqnarray}
as $f^{\lambda \beta \lambda}=0$ on account of the antisymmetric
property of $f$.  For the next-order induced color charge density
element $\rho_{\rm ind}^{a(2)}({\bf r'},{\bf r}) \Delta {\bf r}$, we
can write out explicitly the summations of color and spatial indices
of Eq.\ (\ref{second}),
\begin{eqnarray}
\rho_{\rm ind}^{ \lambda (2)}({\bf r'},{\bf r}) \Delta{\bf r}
=g^2 \sum_{b,c,\beta=1}^8
f^{\lambda b c}    f^{c \beta \lambda}  
\sum_{i,j=1}^3 
\frac{ A^{bi}({\bf r}') ({ r}'^i-{ r}^i) A^{\beta j}({\bf r}) { r}^j }
{|{\bf r}|^3 |{\bf r}' -{\bf r}|^3}   \Delta {\bf r}.
\end{eqnarray}
We note that
\begin{eqnarray}
\sum_{c=1}^8
f^{\lambda b c}    f^{c \beta \lambda}  
=-F(\lambda,b) \delta^{b \beta}, 
\end{eqnarray}
where $F(\lambda,b)$ is a non-negative quantity defined by
\begin{eqnarray}
F(\lambda,b)=\sum_{c=1}^8 (f^{\lambda b c})^2, 
\end{eqnarray}
which can be easily evaluated.  For example, $F(1,b)$ is $\{0,\, 1,\,
1,\, 1/4,\, 1/4,\, 1/4,\, 1/4,\, 0\}$ for $b=1,2,...8$.  In terms of
$F(\lambda, b)$, the induced charge density is
\begin{eqnarray}
\label{AA}
\rho_{\rm ind}^{\lambda (2)}({\bf r}',{\bf r}) \Delta {\bf r} 
=- g^2 \sum_{b=1}^8 F(\lambda, b) \Delta {\bf r}
\sum_{i,j=1}^3  
\frac{ A^{bi}({\bf r}')A^{b j}({\bf r}) ({ r}'^i-{ r}^i) { r}^j }
 { {|{\bf r}|}^3 |{\bf r}' -{\bf r}|^3} .
\end{eqnarray}
\begin{figure} [h]
\includegraphics[angle=0,scale=0.63]{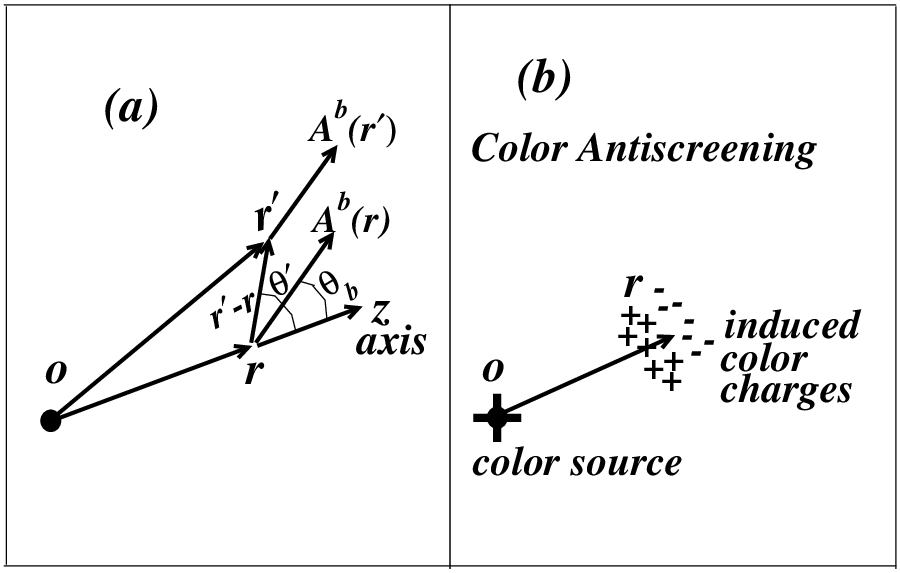}
\caption{($a$) The coordinate system used in the evaluation of the
induced color charge density $\rho_{\rm ind}^{\lambda (2)}({\bf
r}',{\bf r})$ at ${\bf r}'$.  ($b$) A heavy quark of color source at
$O$ induces color charges of the same sign in the direction forward of
${\bf r}$ and color charges of the opposite sign in the direction
backward of ${\bf r}$.  }
\end{figure} 
Note that $F(\lambda, \lambda)=0$.  The contribution of the external
gauge field to the sum over $b$ comes only from those color components
of $A^{b i}$ that are transverse to the color axis $\lambda$ of the
external point source.  

The product $A^{bi} ({\bf r}') A^{bj}({\bf r})$ in Eq.\ (\ref{AA})
involves spatial gauge fields at different spatial locations.
Previously, Svetitsky, Yaffe, DeTar and DeGrand found that large
space-like Wilson loops in the quark-gluon plasma have an area law
behavior.  Such a behavior indicates that the spatial gauge fields
$A^{bj}({\bf r})$ at different spatial locations are correlated
\cite{Yaf82,Sve82,Det85,Deg86,Det88}. 
As shown in Appendix A, if the gauge fields $A^{bi}$ at different field
points ${\bf r}_>$ and ${\bf r}_<$ with $|A^{bi}({\bf r}_>)| > |{ A}^{bi}
({\bf r}_<)|$ are correlated as
\begin{eqnarray}
\label{correl}
{A}^{bi} ({\bf r}_<) ={  A}^{bi}({\bf r}_>) e^{-|{\bf r}_>-{\bf r}_<|/\xi},
\end{eqnarray}
then the integral of ${ A}^{bi} ({\bf r})$ along a space-like loop
$\oint A^{bi} dx^i$ obeys an area law, when the linear dimensions of the
loop are small compared with the correlation length.  Consequently,
within the vicinity of the the correlation length $\xi$, we can
express the thermal average of the $A^{bi} ({\bf r}') A^{bj}({\bf r})$
in terms of the relative coordinate and the correlation
length $\xi$.
\begin{eqnarray}
\langle A^{bi} ({\bf r}') A^{bj}({\bf r}) \rangle \sim A^{bi} ({\bf
 r}) A^{bj}( {\bf r}) e^{-|{\bf r}'-{\bf r}|/\xi}.
\end{eqnarray}
where we have considered the case when $|{ A}^{bi} ({\bf r}')| < |{
A}^{bi} ({\bf r})|$. (The case of $|{ A}^{bi} ({\bf r}')| > |{ A}^{bi}
({\bf r})|$ can be treated in an analogous way).  The quenched QCD
calculations in $SU(3)$ is in the universal class of three-dimensional
$Z(3)$ symmetry.  It has a first order phase transition and and it
possesses a large but finite correlation length at $T_c$
\cite{Yaf82,Sve82,Kog83,Det85,Deg86,Det87,Det88,Fuk89,Hol04,Kar95}.

To study the sign of the induced color charges, we choose a spatial
coordinate system with the $z$-axis along ${\bf r}$ as shown in Fig.\
(4$a$).  In this coordinate system, we label the angular coordinates
of ${\bf r}'-{\bf r}$ and ${\bf A}^{b}({\bf r})$ by $(\theta', \phi')$
and $(\theta_b, \phi_b)$, respectively.  We shall study the induced
charge at ${\bf r}'$ within the vicinity of the correlation length
$\xi$ of ${\bf r}$.  Then, after taking the thermal average, the
induced color charge density is then
\begin{eqnarray}
\rho_{\rm ind}^{\lambda (2)}({\bf r}',{\bf r}) \Delta{\bf r}  
&=& -g^2 \sum_{b=1}^8 F(\lambda,b)
\Delta{\bf r} \, \frac{ 
| {\bf A}^b({\bf r}) |^2
 e^{-|{\bf r}' -{\bf r}|/\xi}}
 {{ r}^2 |{\bf r}' -{\bf r}|^2 }
\nonumber\\
& &~~~~~~~~~~~~~~~~\times  \cos \theta_b [ \cos \theta_b \cos \theta' 
+\sin  \theta_b \sin \theta' \cos (\phi_b -\phi')].
\end{eqnarray}
If we average over the azimuthal angle $\phi'$, the second term in the
square bracket drops out and we have
\begin{eqnarray}
\rho_{\rm ind}^{\lambda (2)}({\bf r}',{\bf r}) \Delta{\bf r}  = 
-g^2 \sum_{b=1}^8  F(\lambda,b) 
\Delta{\bf r} \, 
\frac{| {\bf A}^b ({\bf r})|^2 \,  \cos^2 \theta_b   
~e^{-|{\bf r}' -{\bf r}|/\xi}  \cos \theta' } 
{r^2 (r^2+ |{\bf r}'-{\bf r}|^2+2r|{\bf r}'-{\bf r}|\cos\theta') },
\end{eqnarray}
One readily observes that the induced color charge density $\rho_{\rm
ind}^{\lambda (2)}({\bf r}',{\bf r})$ at ${\bf r}'$ is negative in the
forward hemisphere in the direction forward of ${\bf r}$ (with $\pi/2
\ge \theta'\ge 0$).  It changes to positive in the backward
hemisphere, in the direction backward of ${\bf r}$ ($\pi \ge \theta'
\ge \pi/2$).  In the region of ${\bf r}'$ within the vicinity of the
correlation length from ${\bf r}$, the induced charge surrounding
${\bf r}$ is a color-dipole type density distribution with the color
charge of the same sign at distances closer to the color source and of
the opposite sign at distances farther to the color source (Fig.\
4$b$).  This is the antiscreening behavior due to the presence of
external gauge field $ A^{bi}({\bf r}) $ at ${\bf r}$.  The magnitude
of the induced color charges will increase with an increases in the
correlation length $\xi$ and the magnitude of the gluon field. The
antiscreening effects will enhance the attractive interaction between
the heavy quark and antiquark and will reduce the screening mass from
the Debye screening predictions.

The antiscreening effect arises due to the non-linear properties of
the non-Abelian gauge field while the effects of Debye screening
arises from the interaction between the quark and antiquark with
gluons.  Both effects are present and the antiscreening effects due to
deconfined gluons will act to counterbalance the effects of Debye
screening.  At the onset of the phase transition, the correlation
length is large
\cite{Yaf82,Sve82,Kog83,Det85,Deg86,Det87,Det88,Fuk89,Hol04,Kar95,Yeo92,Lee89}
and deconfined gluons are present, there can be ``remnants of the
confinement part of QCD'' at temperatures slightly above $T_c$, as
pointed out by Kaczmarek $et~al.$ \cite{Kac04}.  At a much higher
temperature, a greater thermal fluctuation leads to a smaller
correlation length, reduces the effects of antiscreening, and Debye
screening dominates.

\section{ Dissociation of $J/\psi$ in collision with gluons} 

It is not necessary to reach the spontaneous dissociation temperature
with zero binding energy for a quarkonium to dissociate.  In a
quark-gluon plasma, gluons and quarks can collide with a color-singlet
heavy quarkonium to lead to the dissociate of the heavy quarkonium.
Dissociation by the absorption of a single gluon is allowed as the
color-octet final state of a free quark and a free antiquark can
propagate in the color medium, in contrast to the situation below
$T_c$ in which the quark and the antiquark are confined.  We shall
consider dissociation of heavy quarkonium by gluons in the present
work.  The collision of a heavy quarkonium with light quarks can also
lead to the dissociation of the heavy quarkonium, but through
higher-order processes.  They can be considered in future refined
treatment of the dissociation process.

Previous treatment of the dissociation of heavy quarkonium by the
absorption of a gluon was carried out by Peskin and Bhanot
\cite{Pes79,Bha79}.  They use the operator product expansion and the large
$N_c$ limit.  They sum over a large set of diagrams and show that to
obtain gauge invariant results, they need to sum over diagrams in
which the external gluon is coupled to the gluon that is exchanged
between the heavy quark and the heavy antiquark.  They use hydrogen
wave function and hydrogen states to evaluate the transition matrix
elements.  Their expression for the dissociation cross section of
$\sigma(g+ (Q\bar Q)_{1S} \to Q + \bar Q)$ is
\begin{eqnarray}
\label{dipole}
\sigma(g+ (Q\bar Q)_{1S} \to Q + \bar Q) = \frac{2}{3}\pi
\left ( \frac{32}{3} \right )^2 
\left ( \frac{4}{3 \alpha_s^2}\right ) 
\frac{1}{m_Q^2} ~\frac{(E/B)^{3/2}}{(E/B+1)^5},
\end{eqnarray}
where $E$ is the non-relativistic kinetic energy of the dissociated
$Q$ and $\bar Q$ in the center-of-mass system. In this short-distance
approach of Peskin and Bhanot, the quark and the antiquark form a
color-dipole pair and the gluons couple to the Wilson loop (the
quarkonium) via simple dipole interactions.  The dissociation cross
section of Eq.\ (\ref{dipole}) is, in fact, the dissociation of the
quarkonium through the absorption of an E1 gluon radiation.

\begin{figure}[h]
\includegraphics[scale=0.50]{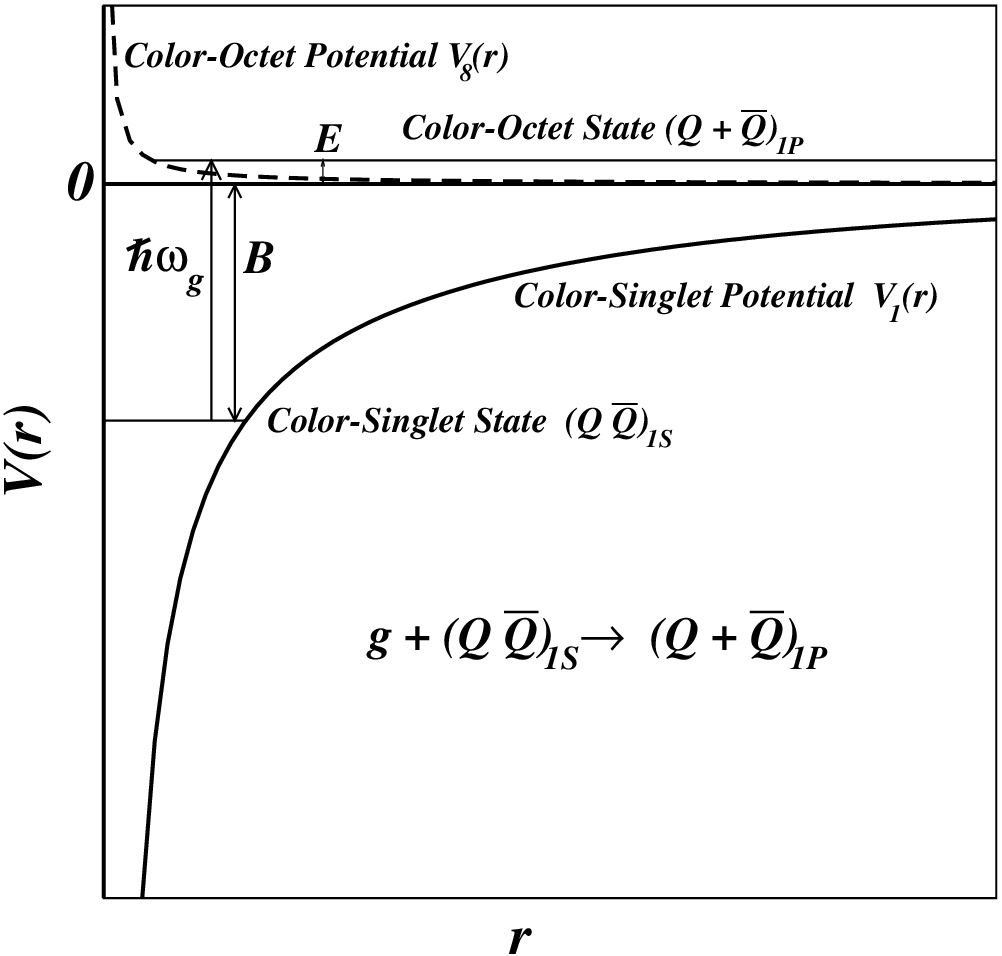}
\caption{The quarkonium dissociation process in the potential model.} 
\end{figure}

Peskin and Bhanot's analytical result for the dissociation cross
section has been applied to many calculations
\cite{Kha94,The01a,The01b}.  In heavy quarkonia of interest, the
radial dependence of the quark-antiquark potential often differ from
the Coulomb potential.  The calculation of the dissociation cross
section requires a new formulation which can be best described by the
potential model introduced previously \cite{Won99,Won02}, following
the results of Blatt and Weisskopf \cite{Bla52} obtained for the
photo-disintegration of a deuteron.  The dissociation process is
schematically illustrated in Fig. 12.  An initial bound $(Q\bar
Q)_{1S}$ state with a binding energy $B$ in the color-singlet
potential $V_1(r)$ absorbs a gluon of energy $\hbar \omega_g$, and is
excited to the color-octet final state $(Q+\bar Q)_{1P}$ with a
kinetic energy $E$ above the rest mass of $m_Q+m_{\bar Q}$.  The
interaction $V_8(r)$ between $Q$ and $\bar Q$ in the color-octet state
will be different from the interaction $V_1(r)$ in the color-singlet
state, as shown in Fig. 12.  At low energies, the dominant dissociation
cross section is the E1 color-electric dipole transition for which the
final state of $Q+\bar Q$ will be in the 1P state in the continuum.
The dissociation cross section $\sigma(g + J/\psi \to c + \bar c)$ for
such a color E1 transition can be obtained from the analogous result
in QED \cite{Won02,Bla52}, and the result is \cite{Won02}
\begin{eqnarray}
\label{pot}
\sigma_{\rm dis}^{E1}(E_{\rm gluon})=4 \times \frac {\pi}{3} 
\alpha_{\rm gQ} (k^2 +\gamma^2)k^{-1}I^2,
\end{eqnarray}
where
\begin{eqnarray}
E_{\rm gluon}=B+E,~~~~~~\gamma^2=2\mu B, 
~~~~~k^2=2\mu E,
\end{eqnarray}
\begin{eqnarray}
I=\int_0^{\infty} u_{1P}(r)~r~ u_{1S}(r) dr ,
\end{eqnarray}
\begin{eqnarray}
\label{alp}
\alpha_{\rm gQ}=\alpha_s |\langle 8c |\frac{\lambda^c}{2} | 1 \rangle |^2 
=\alpha_s  ~\times~ \frac{1}{6} ,
\end{eqnarray}
and $\alpha_s$ is the gluon-(heavy quark) coupling leading to the
dissociation of the heavy quarkonium.  Here, we use the same notation
and normalization as in Blatt and Weisskopf. The bound state wave
function $u_{1S}$ has been normalized according to Eq.\ (\ref{nor}) as
in Eq.\ (XII.4.18) of Blatt and Weisskopf \cite{Bla52}, and the continuum
wave function $u_{1P}$ is normalized according to
\begin{eqnarray}
u_{1P}(r)|_{r \to \infty}  \to kr j_1(kr) = \frac{\sin (kr)}{kr}-\cos (kr), 
\end{eqnarray}
as in Eq.\ (XII.4.32) of Blatt and Weisskopf \cite{Bla52}.  The result from
the potential model agrees with the analytical results of Bhanot and
Peskin for the case they considered (hydrogen wave function, large
$N_c$ limit,...) \cite{Won02}.  Such an agreement was further
confirmed by numerical calculations according to Eq.\ (\ref{pot})
using hydrogen wave function for $u_{1S}$ and plane wave continuum
wave function for $u_{1P}$, as assumed by Peskin and Bhanot
\cite{Pes79,Bha79}.  The potential model has the practical advantage
that it can be used for a $Q$-$\bar Q$ system with a general
potential.

\section{$J/\psi$ Collisional Dissociation Rate and 
Dissociation Width}

We have represented the color-singlet potential between a heavy quark
and antiquark by a screened Coulomb potential and have obtained the
$J/\psi$ wave function. To study the gluon dissociation of $J/\psi$,
we need the color-octet potential $V_8({\bf r},T)=U_{Q\bar
Q}^{(8)}({\bf r},T)$ experienced by the $Q$ and $\bar Q$ in the final
state.  We shall assume the generalization that the color-dependence
of the potential Eq.\ (\ref{scoulF}) is simply obtained by modifying
the color factor from $-4/3$ for the color-singlet state to $1/6$ for
the color-octet state:
\begin{eqnarray}
U_{Q\bar Q}^{(i)}({\bf r},T)-U_{Q\bar Q}^{(i)}({\bf r}\to\infty,T) &=&
C_f \frac {f_F \alpha_F~e^{-\mu_F~ r} +f_U\alpha_U~e^{-\mu_U~ r} }
{r}, \\ C_f &=&
\begin{cases}
   -4/3 ~~(i=1,{\rm~color-singlet})\cr 
  ~~1/6 ~~(i=2,{\rm~color-octet}). \cr   
\end{cases}
\end{eqnarray}
We also need the gluon-quark coupling constant $\alpha_s$ in Eq.\
(\ref{alp}) to evaluate the dissociation cross section.  We shall
consider the screened Coulomb potential (\ref{pot}) obtained in
the lattice gauge theory as arising from the exchange of a virtual
nonperturbative gluon and assume that the coupling of gluon to the
heavy quark leading to quarkonium bound states,
$f_F\alpha_F+f_U\alpha_U$, is the same coupling leading to the
dissociation of the quarkonium.  The $J/\psi$ dissociation cross
section can then be calculated using Eq. (\ref{pot}).  The results of
the $J/\psi$ dissociation cross section for different temperatures are
shown in Fig. 13.  The cross section increases up to a maximum value
and decreases as the gluon energy increases.  As the temperature
decreases the maximum height of the dissociation cross section
increases, but the width of the cross section decreases.  We shall
limit our attention to the dissociation of $J/\psi$ by the absorption
of an $E1$ radiation in the present analysis.  When the dissociation
threshold decreases, higher multipole dissociation may become
important.  It will be of interest to study dissociation arising from
gluon radiation of higher multipolarity in future work.

\begin{figure}[h]
\includegraphics[scale=0.45]{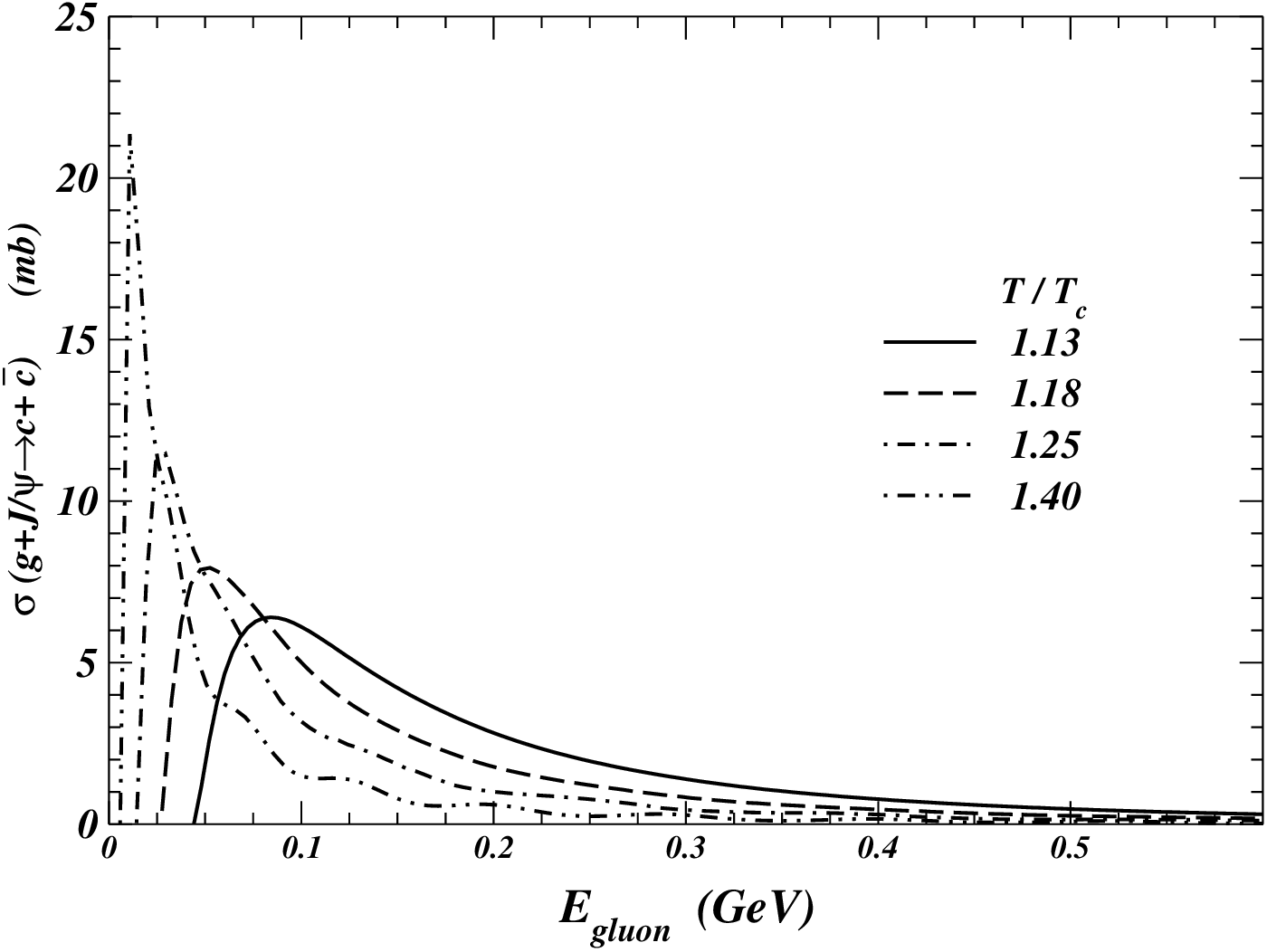}
\caption{$J/\psi$ dissociation cross section as a function of gluon
energy at various quark-gluon plasma temperatures.}
\end{figure}

If we place a $J/\psi$ in a quark-gluon plasma at a temperature $T$,
the average $E1$ dissociation cross section is
\begin{eqnarray}
\langle \sigma_{\rm dis}^{E1}\rangle = \frac{g_g}{2\pi^2}
\int_0^\infty
\sigma_{\rm dis}^{E1}(p) \frac { p^2 \, dp } {e^{p/T}-1} \biggr / n_g,
\end{eqnarray}
where $g_g=16$ is the gluon degeneracy and $n_g$ is the gluon density,
\begin{eqnarray}
n_g 
= \frac{g_g}{2\pi^2}
\int_0^\infty
\frac { p^2 \, dp } {e^{p/T}-1}. 
\end{eqnarray}
Using the energy dependence of the dissociation cross section as given
in Fig.\ 13, we can calculate the average dissociation cross section
$\langle \sigma_{\rm dis}^{\rm E1} \rangle $.  From $\langle
\sigma_{\rm dis}^{\rm E1} \rangle $, we obtain the rate of $J/\psi$
dissociation (by E1 transition) given by
\begin{eqnarray}
\frac {dn_{J/\psi}} {dt} = - n_g \langle \sigma_{\rm dis}^{\rm E1} \rangle .
\end{eqnarray}
This dissociation rate leads to the collisional dissociation width
$\Gamma_{\rm E1}$ due to the absorption of $E1$ gluon radiation,
\begin{eqnarray}
\Gamma_{\rm E1}=n_g
\langle \sigma_{\rm dis}^{\rm E1}\rangle.
\end{eqnarray}

We show in Fig. 14 the temperature dependence of $\langle \sigma_{\rm
dis}^{\rm E1} \rangle $ and $\Gamma_{\rm E1}$.  One observes that the
average cross section is in the range of 0.2-0.8 mb.  The collisional
dissociation width due to $E1$ gluon absorption is of the order of
0.05-0.11 GeV, and the mean life of $J/\psi$ in the quark-gluon plasma
due to the absorption of gluons to the 1P state is therefore of order
2-4 fm/c.

\begin{figure} [h]
\includegraphics[scale=0.50]{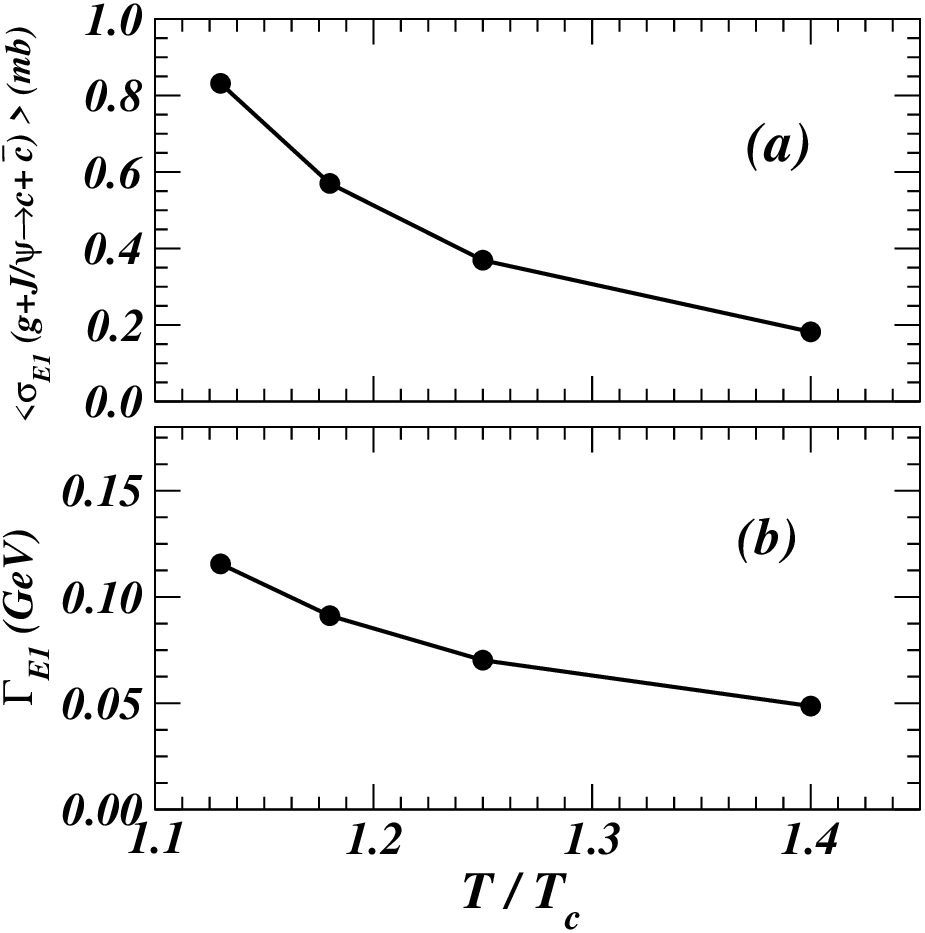}
\caption{(a) Thermally averaged dissociation cross section as a
function of temperature. (b) $J/\psi$ collisional dissociation width
$\Gamma_{E1}$ as a function of temperature.}
\end{figure}

\section{$J/\psi$ Production by the collision of $c$ and $\bar c$ in QGP}

In high-energy nuclear collisions, elementary nucleon-nucleon
collisions lead to the production of open heavy quark mesons.
Although the probability for such a production is small for a single
nucleon-nucleon collision, there are many nucleon-nucleon collisions
in a central nucleus-nucleus collision. The large number of binary
nucleon-nucleon collisions can produce many pairs of open charm
mesons, and these open charm mesons can recombine to produce $J/\psi$.
It is of interest to estimate the elementary reaction cross section
for $c+ \bar c \to J/\psi +g$ and obtain the rate of $J/\psi$
production in a nucleus-nucleus collision.

The reaction $c+\bar c\to J/\psi +g$ is
just the inverse of $g+ J/\psi \to c+\bar c$.  Their cross sections
are therefore related by \cite{Bla52a}
\begin{eqnarray}
\label{invs}
\sigma^{E1}(c+ \bar c\to J/\psi +g)
=
\frac {|\bbox{p}_1|^2}{|\bbox{p}_3|^2}
\sigma^{E1}(J/\psi +g \to c+ \bar c ),
\end{eqnarray}
where $\bbox{p}_1$ is the momentum of one of the particles in the
$J/\psi +g$ system, and $\bbox{p}_3$ is the momentum of one of the
particles in the $c +\bar c$ system, both measured in the
center-of-mass frame.  With Eq.\ (\ref{invs}) and the results of
$\sigma^{E1}(J/\psi +g \to c+ \bar c )$ in Fig.\ 13, the production
cross section $\sigma^{E1}(c+ \bar c\to J/\psi +g)$ can be calculated.
The cross section as a function of the kinetic energy of $c$ and $\bar
c$ in the center-of-mass system are shown in Fig. 15.  One observes
that the cross section peaks at low kinetic energies near the
threshold, and the magnitude of the cross section decreases with
temperature.  The maximum cross section at $T/T_c \sim 1.13$ is of
order 0.7 mb.

\begin{figure} [h]
\includegraphics[scale=0.45]{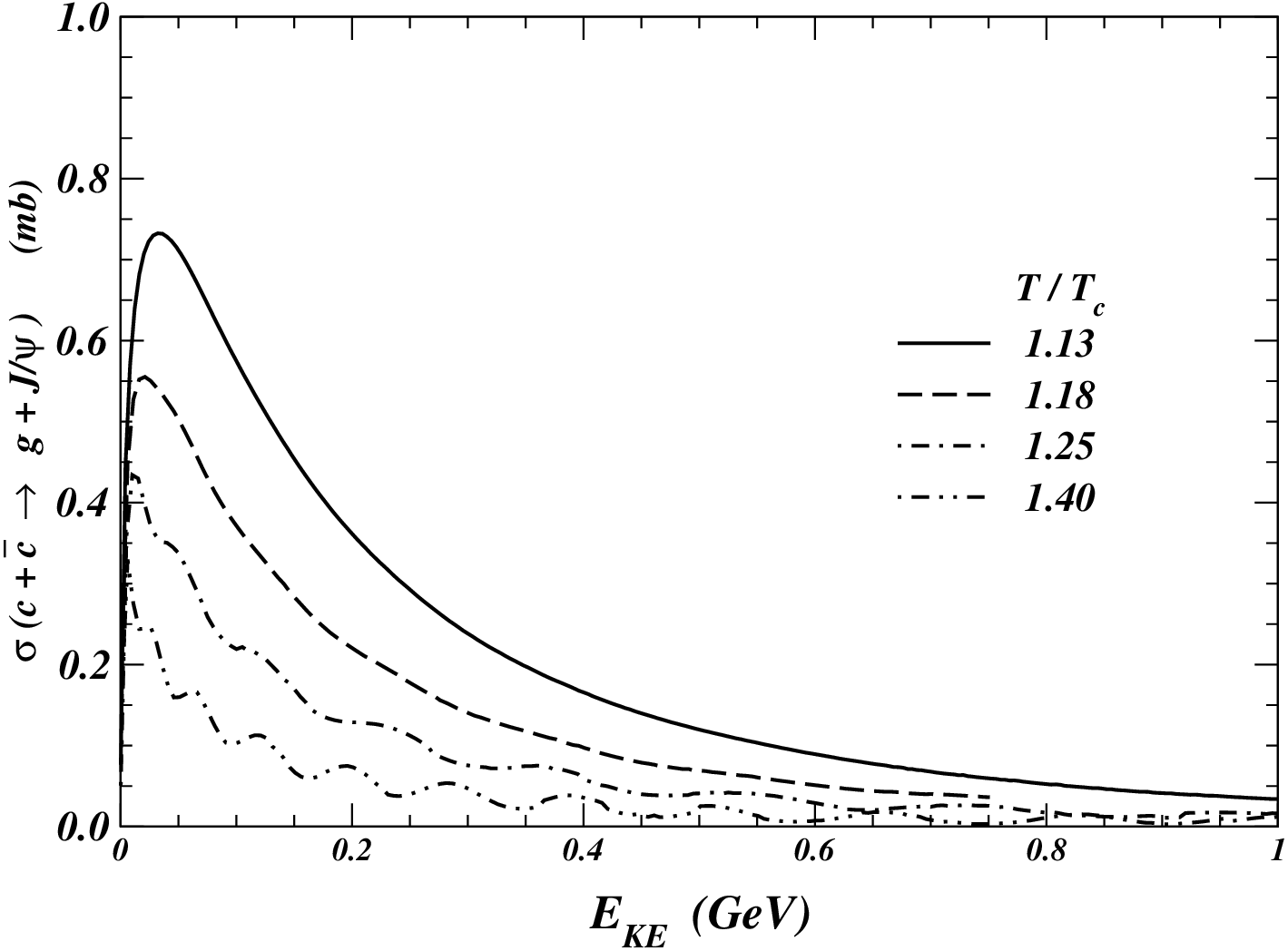}
\caption{Cross section for the production of
$J/\psi$ by the collision of $c$ and $\bar c$}
\end{figure}

The rate of $J/\psi$ production can be obtained when the momentum
distribution $f(y,\bbox{p}_t)$ of the produced $c$ and $\bar c$ is
known. For simplicity, we consider charm quarks and antiquarks to be
contained in a spatial volume $V$ with a uniform distribution. The
probability of producing a $J/\psi$ in the volume $V$ per unit time by
the collision of charm quark and antiquark is $\sigma(c+\bar c \to
J/\psi+g) v_{12}/V$, where $v_{12}$ is the relative velocity between
$c$ and $\bar c$.  The number of $J/\psi$ produced per unit time from
collision of $c$ and $\bar c$ is therefore
\begin{eqnarray}
\label{prod}
\frac {dN_{J/\psi}} {dt} =\int dy_1 d\bbox{p}_{1t}\, dy_2 d\bbox{p}_{2t}
\, f(y_1,\bbox{p}_{1t})\, f(y_2,\bbox{p}_{2t}) 
\, \sigma(c +\bar c \to J/\psi+g) v_{12}/V.
\end{eqnarray}
When we neglect initial- and final-state interactions, the momentum
distribution of charm is given by
\begin{eqnarray}
f(y,\bbox{p}_t)=N_{\rm bin}\frac{E dN_{c\bar c}^{\rm pp}}{d\bbox{p}}.
\end{eqnarray}
Here $f(y,\bbox{p}_t)$ is normalized to the total number of charm
quarks $N_c$ and antiquarks $N_{\bar c}$ produced in the
nucleus-nucleus collision, $N_c=N_{\bar c}=\int (d\bbox{p}/E)
f(y,\bbox{p}_t)=N_{\rm bin} N_{c\bar c}^{\rm pp}$, $N_{\rm bin}$ is
the number of binary nucleon-nucleon collisions, and $N_{c\bar c}^{\rm
pp}$ is the number of $c\bar c$ pair produced in a single
nucleon-nucleon collision.  From the charm production data in $d$-Au
collisions and PYTHIA calculations, Tai $et~al.$ \cite{Tai04}, Adams
$et~al.$ \cite{Ada04}, and the STAR Collaboration inferred that at
$\sqrt{s}=200$ GeV the charm production cross section per
nucleon-nucleon collision is $\sigma_{c\bar c}^{\rm pp}|_{\rm
STAR}=1.18\pm 0.21($stat$)\pm 0.39($sys) mb \cite{Tai04} and
$\sigma_{c\bar c}^{\rm pp}|_{\rm STAR}=1.4 \pm 0.2\pm 0.4 $ mb
\cite{Ada04}.  If one uses the transverse momentum distribution
measured and parametrized by Tai $et~al.$ \cite{Tai04} and assumes a
Gaussian rapidity distribution, the charm momentum distribution of
Adams $et~al.$ \cite{Ada04} per nucleon-nucleon collision can be
represented by
\begin{eqnarray}
\label{data2}
\frac { E dN_{c\bar c}^{\rm pp} } { d\bbox{p} }({\rm PYTHIA(STAR)})=
\frac { E d \sigma_{c \bar c}^{pp}|_{\rm STAR} } { \sigma_{in} d \bbox{p} }
=A\frac {e^{ -y^2/2 \sigma_y^2 } }  {( 1+p_t/p_{t0} )^n},
\end{eqnarray}
where $\sigma_{in}=42$ mb is the nucleon-nucleon inelastic cross
section, $A=4.4\times 10^{-3}$ (GeV$^{-2})$, $\sigma_y=1.84$,
$p_{t0}=3.5$ GeV, and $n=8.3$.  The number of $c\bar c$ produced per
nucleon-nucleon collision is $N_{c \bar c}^{\rm pp}|_{\rm STAR}=1.4$ mb/42
mb=0.033$\pm 0.0107$.  

The PHENIX Collaboration obtained $\sigma_{c\bar c}^{\rm pp}|_{\rm
PHENIX}=622\pm 57({\rm stat})\pm160 ({\rm sys})~\mu$b \cite{Adl04} for
the open charm production cross section per nucleon-nucleon collision
at $\sqrt{s}=200$ GeV, and $N_{c \bar c}^{\rm pp}|_{\rm
PHENIX}=0.0148\pm 0.004$.  With this total charm production cross
section, the theoretical results from the PYTHIA calculations of the
PHENIX Collaboration can be parametrized as \cite{Adl04,Sil04}
\begin{eqnarray}
\label{data3}
\frac { E dN_{c\bar c}^{\rm pp} } { d\bbox{p} }({\rm PYTHIA(PHENIX)})
=A'\frac {e^{ -y^2/2 \sigma_y'^2 } }  {( 1+p_t/p_{t0}' )^{n'}},
\end{eqnarray}
where  $A'=6.48\times 10^{-4}$ (GeV$^{-2})$, $\sigma_y'=1.85$,
$p_{t0}'=5.06$ GeV, and $n'=7.0$. 

We shall focus attention to central Au-Au collisions within the most
inelastic 10\% of the reaction cross section.  The average number of
binary collision $N_{\rm bin}$ for these central Au-Au collisions at
RHIC is $N_{\rm bin}=833$. For these central nucleus-nucleus
collisions at $\sqrt{s}=200$ GeV, the average number of $c$ and $\bar
c$ produced is $N_c=N_{\bar c}=27.8\pm 8.9$ if we use the cross
section of the STAR Collaboration \cite{Ada04}, and $N_c=N_{ \bar
c}=12.34 \pm 3.4 $ if we use the cross section of the PHENIX
Collaboration \cite{Adl04}.  The rate of $J/\psi$ production can then
be obtained from Eqs.\ (\ref{prod})-(\ref{data3}) by carrying out the
six-fold integration.

In Fig. 16 we show the quantity $V dN_{J/\psi}/dt$ as a function of
temperature for the most inelastic (10\%) central Au-Au collisions at
$\sqrt{s}=200$ GeV.  The estimate of the rate of $J/\psi$ production,
using the momentum distribution of Eq.\ (\ref{data2}) from the PYTHIA
calculations of the STAR Collaboration \cite{Ada04}, is greater than
the corresponding estimate, using the momentum distribution of Eq.\
(\ref{data3}) from the PYTHIA calculations of the PHENIX Collaboration
\cite{Ada04,Sil04}, by a factor of about 10.  This factor is larger
than the factor of 2.25 of the nucleon-nucleon $c \bar c$ production
cross section of the STAR Collaboration, relative to the corresponding
cross section of the PHENIX Collaboration at $\sqrt{s}=200$ GeV.  The
large difference of these two factors arises because the charm
momentum distribution from the PYTHIA calculations of the STAR
Collaboration has a greater magnitude at low $p_T$, and $c$ and $\bar
c$ recombine more readily at low relative energies, as indicated in
Fig.\ 16.

\hspace*{2cm}
\begin{figure}[h]
\includegraphics[scale=0.55]{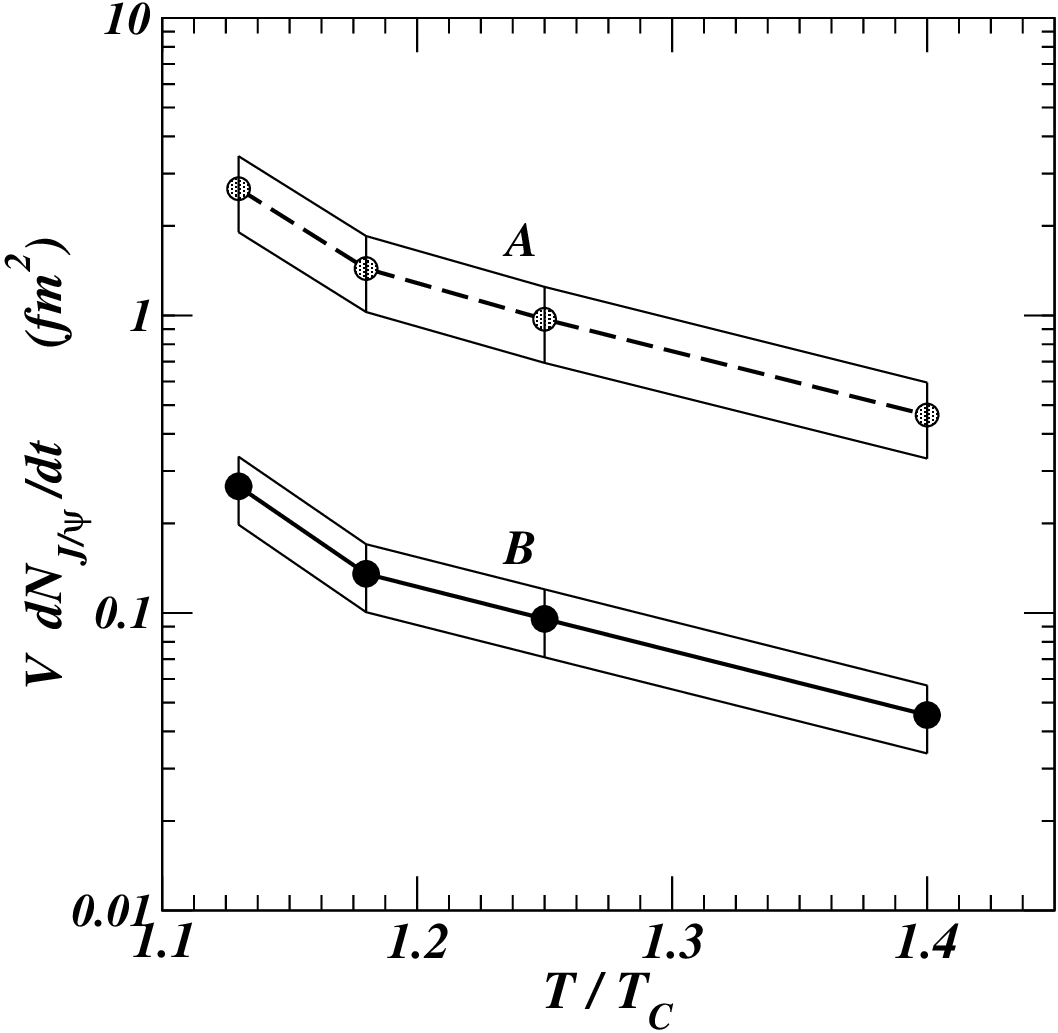}
\caption{The rate of $J/\psi$ production as a function of temperature,
for the most inelastic (10\%) central Au-Au collision at
$\sqrt{s}=200$ GeV.  Curve A is based on the charm momentum
distribution of Eq.\ (\ref{data2}) using the PYTHIA calculations of
the STAR Collaboration \cite{Ada04}, and Curve B is based on the charm
momentum distribution of Eq.\ (\ref{data3}) using the PYTHIA
calculations of the PHENIX Collaboration \cite{Ada04,Sil04}.}
\end{figure}

We can illustrate the magnitude of the rate of $J/\psi$ production by
considering a Au-Au central collision with a transverse area of $\pi
(7 {\rm~fm})^2$ and a longitudinal initial time (and longitudinal
length) of 1 fm.  The initial volume containing the charm quarks and
antiquarks is about 150 fm$^3$.  If the initial temperature is
1.4$T_c$, the initial rate of $J/\psi$ production will be about $3\times10^{-4}$
to $3\times10^{-3}$ fm/c.  As the volume expands, the temperature
decreases.  The results of Fig. 17 can be use to provide an estimate
of the rate of $J/\psi$ production.

\section{Discussions and Conclusions}

We use the color-singlet free energy $F_1$ and internal energy $U_1$
obtained by Kaczmarek $et~al.$ \cite{Kac03} in quenched QCD to study
the energy levels of charmonium and bottomium above the phase
transition temperature.  From a variational principle in a schematic
model, we find the the $Q$-$\bar Q$ potential involves only the
$Q$-$\bar Q$ internal energy $U_{Q\bar Q}^{(1)}$, which can be
obtained from the total $U_1$ by subtracting the gluon internal energy
contributions.  We carry out this subtraction using the local
energy-density approximation in which the gluon energy density can be
related to the local gluon pressure by the quark-gluon plasma equation
of state.  We find that the $Q$-$\bar Q$ potential is $U_{Q\bar
Q}^{(1)}=3F_1/(3+a)+aU_1/(3+a)$ where $a=3p/\epsilon$ is given by the
quark-gluon plasma equation of state.  Such a $U_{Q\bar Q}^{(1)}$
potential leads to weakly bound $J/\psi$ and $\eta_c$ at temperatures
above the phase transition temperature and they become unbound at
$1.62 T_c$.  The $\chi_c$, $\eta_c'$ and $\psi'$ states are found to
be unbound in the quark-gluon plasma.  In this potential model,
$\Upsilon$, $\eta_b$, $\Upsilon'$, $\eta_b'$ and $\chi_b$ are bound at
temperatures above $T_c$ and $\Upsilon$ and $\eta_b$ dissociates
spontaneously at $4.1 T_c$, $\chi_b$ at $1.18 T_c$ and $\Upsilon'$,
and $\eta_b'$ at $1.38 T_c$.

The results from the $U_{Q\bar Q}^{(1)}$ potential need to be tested
against results from spectral function analysis.  For completeness, we
have also calculated heavy quarkonium binding energies using the free
energy $F_1$ \cite{Dig01a,Dig01b,Won01a,Won02} and the internal energy
$U_1$ as the potential \cite{Kac02,Kar03,Zan03}.

The comparison shows that different models give very different heavy
quarkonium binding energies.  The potential that agrees best with
results obtained from spectral function analysis is the $U_{Q\bar
Q}^{(1)}$ potential deduced in the present analysis.  The agreement
with spectral function analysis and the theoretical foundations
presented here provide support for the use of $U_{Q\bar Q}^{(1)}$ as
the proper $Q$-$\bar Q$ potential in heavy quarkonium studies.
Conversely, the agreement also lends support to the quantitative
features concerning the stability of heavy quarkonia in the spectral
function analyses of Asakawa $et~al.$ \cite{Asa03,Asa04} and Petreczky
$et~al.$ \cite{Pet03,Dat03,Pet04}.

The spectral function analysis for the bottomium states has not yet
been carried out.  As the predications for the dissociation
temperatures for bottomium states are quite different from different
potential models, it will be of great interest to calculate the
bottomium dissociation temperatures in lattice gauge spectral function
analysis so as to test the potential models further.

In a nucleus-nucleus collision, charm quarks and antiquarks are
produced in hard-scattering processes in nucleon-nucleon collisions.
During the time of a central nuclear reaction, these heavy quarks 
and antiquarks will be present in the quark-gluon plasma and can
interact to form $J/\psi$.  We have calculated the cross section for
$J/\psi$ production by the collision of charm quark and antiquark. The
cross section is energy dependent, and the maximum cross section
increases as the temperature decreases.  The production cross section
can be used to study the rate of $J/\psi$ production in
nucleus-nucleus collisions.

We have carried out the investigation using the quenched QCD.  It will
be of interest to carry out similar investigations using unquenched
QCD.  Results of the full QCD in two flavors \cite{Kac03b} and in three
flavors \cite{Pet04a} have been obtained recently, and an
investigation on $J/\psi$ dissociation temperatures in QCD with two
flavors has been initiated \cite{Shu04a}.  A thorough study of how the
dynamical quarks will affect the stability, the dissociation, and the
inverse production of heavy quarkonium will be of great interest.

It is necessary to emphasize that the present $Q$-$\bar Q$ potential
$U_{Q\bar Q}^{(1)}$ of Eq.\ (\ref{uqqq}) extracted from $F_1$ and
$U_1$ has been obtained in the local energy-density approximation.  It
would be of great interest in future lattice gauge calculations to
evaluate the $U_{Q\bar Q}^{(1)}({\bf r})$ directly to check the
validity of the local energy-density approximation.

The color-singlet correlator of the Polyakov lines in Eq.\
(\ref{free1}) is not gauge invariant.  It has been suggested that one
can dress the Polyakov lines to make a gauge invariant definition
of the color singlet potential.  The dressing of the source may be
viewed as a gauge transformation and is equivalent to the choice of a
certain gauge \cite{Phi02} with the requirement that the gauge-fixed
Polyakov loop correlation function in the singlet channel falls off
with gauge-invariant eigenvalues of the Hamiltonian.  This requirement
may be satisfied for the Coulomb gauge and other time-local gauges.
Recent calculations by Belavin $et~al.$
\cite{Bel04} show however that the color singlet potential depends on
the choice of the gauges even among these time-local gauges.  It has
been found that at finite temperatures all channels receive
contributions only from the color-singlet channel.  The extraction of
the color-singlet potential from the "color-singlet" Polyakov
correlator of Eq.\ (\ref{free1}) may include additional $r$ and/or $T$
dependence which is not shared by the physical states \cite{Jah04}.
Clearly, much work remains to be carried out to clarify the proper
color-singlet potential in lattice gauge calculations.

\begin{acknowledgments}
The authors would like to thank Drs. H. Crater, P. Petreczky,
M. Gyulassy, Su Houng Lee, Keh-Fei Liu, T. Barnes, S. Ohta,
V. Cianciolo, D. Silvermyr, Huan Huang, and Zhangbu Xu for helpful
discussions and communications.  This research was supported in part
by the Division of Nuclear Physics, U.S. Department of Energy, under
Contract No. DE-AC05-00OR22725, managed by UT-Battelle, LLC and by the
National Science Foundation under contract NSF-Phy-0244786 at the
University of Tennessee.
\end{acknowledgments}

\appendix
\section{Space-Like Area Law and the Correlation of Gauge Fields} 

We focus our attention on the $x-y$ plane so that the $z$ coordinate
can be omitted and consider a loop integral $\oint_L A_i dx^i$ along
the loop $L$ defined by $(0,0) \to (L_x,0) \to (L_x,L_y) \to (0,L_y)
\to (0,0)$.  The integral around this loop of area $L_x L_y$ is given
approximately by
\begin{eqnarray}
\oint_L  A_i  dx^i = A_x(\frac{L_x}{2},0)L_x +A_y(L_x,\frac{L_y}{2})L_y
- A_x(\frac{L_x}{2},L_y)L_x - A_y(0,\frac{L_y}{2})L_y.
\end{eqnarray}
We can write the right hand side in the form
\begin{eqnarray}
\label{loop}
\oint_L  A_i  dx^i = c L_x L_y,
\end{eqnarray}
where
\begin{eqnarray}
\label{a1}
c= \frac{A_x(L_x/2,0)    -A_x(L_x/2,L_y)}{L_y}
  -\frac{A_y(L_x,  L_y/2)-A_y(0,  L_y/2)}{L_x}.
\end{eqnarray}
If the gauge fields ${\bf A}$ at different field points are
correlated by a correlation length $\xi$ such that for two points
${\bf r}_>$ and ${\bf r}_<$ where $|A_i({\bf r}_>)| > |{ A}_i ({\bf
r}_<)|$ and
\begin{eqnarray}
\label{correla}
{A}_i ({\bf r}_<) ={  A}_i ({\bf r}_>) e^{-|{\bf r}_>-{\bf r}_<|/\xi},
\end{eqnarray}
then we have
\begin{eqnarray}
\frac{A_x(L_x/2,0)    -A_x(L_x/2,L_y)}{L_y}
=\begin{cases}
({1-e^{-L_y/\xi}}) A_x(\frac {L_x}{2},0)/L_y 
{\rm~~~ if ~~~~} A_x(\frac{L_x}{2},0) > A_x(\frac{L_x}{2},L_y), \cr
({e^{-L_y/\xi}}-1) A_x(\frac {L_x}{2},L_y)/L_y 
{\rm~~~ if ~~~~} A_x(\frac{L_x}{2},0) < A_x(\frac{L_x}{2},L_y). \cr
\end{cases}
\end{eqnarray}
The second term in Eq.\ (\ref{a1}) can be similarly evaluated.  In the
case of a correlation length $\xi$ that is large compared with the loop
lengths $L_x$ and $L_y$, the quantity $c$ in Eq.\ (\ref{loop}) can be
evaluated and we obtain the area law
\begin{eqnarray}
\label{area}
\oint_L  A_i  dx^i = \frac{1}{\xi}(A_x^> - A_y^>)  L_x L_y,
\end{eqnarray}
where
\begin{eqnarray}
A_x^>
=\begin{cases}
A_x( {L_x}/{2},0) 
{\rm~~~ if ~~~~} A_x({L_x}/{2},0) > A_x( {L_x}/{2},L_y), \cr
- A_x( {L_x}/{2},L_y)
{\rm~~~ if ~~~~} A_x({L_x}/{2},0) < A_x( {L_x}/{2},L_y), \cr
\end{cases}
\end{eqnarray}
and
\begin{eqnarray}
A_y^>
=\begin{cases}
A_y( 0, {L_y}/{2}) 
{\rm~~~ if ~~~~} A_y(0, {L_y}/{2}) > A_y( {L_x},L_y/2), \cr
- A_y( 0,{L_y}) 
{\rm~~~ if ~~~~} A_y(0, {L_y}/{2}) < A_y( {L_x},L_y/2). \cr
\end{cases}
\end{eqnarray}
Eq. (\ref{area}) shows that if space-like gauge fields at different
points are correlated by Eq.\ (\ref{correla}) with a large correlation
length, the integral of the gauge fields along a space-like Polyakov
loop will satisfy an area law.

\end{document}